\documentclass[final]{svjour3}

\usepackage{graphicx,rotating,epsfig}
\usepackage{amssymb}
\usepackage{mathptmx}
\usepackage[numbers]{natbib}
\usepackage{amsfonts}
\usepackage{amsmath}
\usepackage{amssymb}
\usepackage{color}

\makeatletter
\journalname{Journal of Low Temperature Physics}

\newcommand{\HeThree}{$^3$He}
\newcommand{\HeFour}{$^4$He}

\begin{document}

\title{Defects and glassy dynamics in solid \HeFour: Perspectives and current status}

\author{ A. V. Balatsky$^{1,2}$, M. J. Graf$^{1}$, Z. Nussinov${^3}$, J.-J. Su$^{4}$ }

\institute{1: Theoretical Division, Los Alamos National Laboratory, Los Alamos, New Mexico 87545, USA
\\2: Center for Integrated Nanotechnologies, Los Alamos National Laboratory, Los Alamos, New Mexico 87545, USA
\\3: Department of Physics, Washington University, St.\ Louis, Missouri 63160, USA
\\4: Department of Electrophysics, National Chiao Tung University, Hsinchu 30013,
Taiwan}

\date{09.04.2012}
\maketitle

\begin{abstract}
We review the anomalous behavior of solid \HeFour\ at low temperatures with particular attention to the role of structural defects present in solid. The discussion centers around the possible role of two level systems and structural glassy components for inducing the observed anomalies. We propose that the origin of glassy behavior is due to the dynamics of defects like dislocations formed in \HeFour. Within the developed framework of glassy components in a solid, we give a summary of the results and predictions for the effects that cover the  mechanical, thermodynamic, viscoelastic, and electro-elastic contributions of the glassy response of solid \HeFour.
Our proposed glass model for solid \HeFour\ has several implications:
(1) The anomalous properties of \HeFour\ can be accounted for by allowing defects to  freeze out at lowest temperatures. The dynamics of solid \HeFour\ is governed
by glasslike (glassy) relaxation processes and the distribution of relaxation times varies significantly between
different torsional oscillator, shear modulus, and dielectric function experiments.
(2) Any defect freeze-out will be accompanied by thermodynamic signatures consistent with entropy contributions from defects.  It follows that such entropy contribution is much smaller than the required superfluid fraction, yet it is sufficient to account for excess entropy at lowest temperatures.
(3) We predict a Cole-Cole type relation between the real and imaginary part of the response functions for rotational and planar shear that is occurring due to the dynamics of defects. Similar results apply for other response functions.
(4) Using the framework of glassy dynamics, we predict low-frequency yet to be measured electro-elastic features in defect rich \HeFour\ crystals. These  predictions allow one to directly  test the ideas and very presence of glassy contributions in  \HeFour.

\end{abstract}



\section{Introduction}

The discovery of anomalous frequency and dissipation behavior seen in torsional oscillators (TOs) \cite{Kim04a,Kim04b}
at low temperatures in \HeFour\ has stimulated numerous investigations. These anomalies have been argued to
demonstrate a non-classical rotational inertia (NCRI)
of the long ago predicted supersolid quantum state\cite{Andreev69,Chester67,Reatto69,Leggett70,Anderson84}.
Successive TO experiments \cite{Rittner06,Kondo07,Aoki07,Clark07,Penzev07,Hunt09,Pratt11,Gadagkar2012}
confirmed the finding of the reported anomalous behavior.
Hysteresis behavior and long equilibration times have been observed
\cite{Aoki07,Hunt09,Kim09},
which depend strongly on growth history and annealing \cite{Rittner06}. In the same temperature range,  experiments including shear modulus \cite{Beamish05,Beamish06}, ultrasonic \cite{Goodkind02,Burns93} and heat propagation \cite{Goodkind02} have also shown various anomalies. The character and the existence of mass flow is still a matter of
intense investigation. Experiments designed to probe for mass flow by squeezing the lattice report  no such flow \cite{Beamish05,Beamish06,Greywall77,Paalanen81,Sasaki06,Ray08,Bonfait89,Balibar08}.
However, experiments in which a chemical potential gradient was created via coupling to a superfluid reservoir, suggest mass flow of an unusual type \cite{Ray08,Ray09,Ray10,Ray11,Vekhov2012}. These and many other results have
led to a flurry of activity. Structural measurements \cite{Burns08,Blackburn07} suggest
that solid \HeFour\ may be composed of a dynamic mosaic of crystals, highlight the importance of defects, and report the absence of notable structural change in the vicinity of the putative supersolid transition.
The intricate structure and dynamics of sold \HeFour\ make it a fascinating system.
We note that when a mosaic of \HeFour\  crystals with intervening liquid channels is placed in a metallic container
\cite{Ray08,BalibarComment}, thermal expansion effects
(in particular the larger thermal expansion coefficient of helium vis a vis that of the metallic container) may be at play in blocking a remnant
superleak. Such a thermal compression will effectively shut down any mass flow and close the open channels as the temperature is raised. This scenario remains a viable explanation for the anomalous mass flow and fountain effect reported by Hallock's group \cite{Ray08,Ray09,Ray10,Ray11,Vekhov2012}, until a superleak can be ruled out.

After eight years of intense experimental and theoretical investigations  one must ask what are the established facts, what are the current expectations and hypotheses, and what are the future directions of research in solid \HeFour. There are numerous reviews and progress updates available describing the field
\cite{ProkofevAdvances,Balibar2011,Boninsegni2012}
and addressing some of these questions. Most of the literature reviewing the subject deals with the notion of supersolidity in \HeFour\ as established and proceeds with the discussion on the status and future experiments with the goal of further ``proving" the existence of a supersolid phase transition at very low temperatures.
Amongst the many exciting proposals concerning \HeFour\ as well as supersolids, we briefly mention Andreev's proposal for superglass \cite{Andreev07, Andreev09, Korshunov09},  Anderson's suggestion of vortex proliferation and flow \cite{Anderson2007,Kubota} and supersolid dislocation cores
\cite{dislocation_core1,dislocation_core2,dislocation_core3,Rossi}. As will become evident in later sections, our analysis centers on the dynamical effects of defects and as such may include
vortices, dislocations, or any other defects. Estimating the product of typical dislocation core sizes ($\sim 3$ nm) and their density ($\sim 10^{10}$ m$^{-2}$), it is seen that a direct NCRI origin from superfluid dislocation cores in \HeFour\ cannot account for the
magnitude of the TO anomaly (it is orders of magnitude too small).

We believe that at this mature stage of research in \HeFour\ a somewhat more general  view on possible options is needed, where one asks the question: what are the options for possible states that might  form at low temperatures in \HeFour\ and by implication in other solid bosonic matter? By taking a broader view one  explicitly allows for states other than pure supersolidity and includes possible coexistence phases to form as well.  This review is a contribution in the spirit of broadening the conversation by explicitly allowing for other components or ``active ingredients" to be present in addition to, or perhaps instead of, supersolidity in \HeFour. Thus, if one accepts the presence of defects in a solid, then naturally the question arises about the dynamic signatures of such crystal defects and whether they can dominate the response to an external stimulus.

\begin{figure}
\begin{center}
\parbox[t]{0.4\linewidth}{
\caption{The phase diagram of \HeFour\ with the putative supersolid phase transition below 0.3 K \cite{Kim04b} or an alternate order-disorder dominated crossover region governed by impedance matching between the applied frequency and internal relaxation processes.
}\label{fig_phasediagram}
}
\hfill
\begin{minipage}{0.45\linewidth}
\includegraphics[clip,width=1.0\linewidth,angle=0]{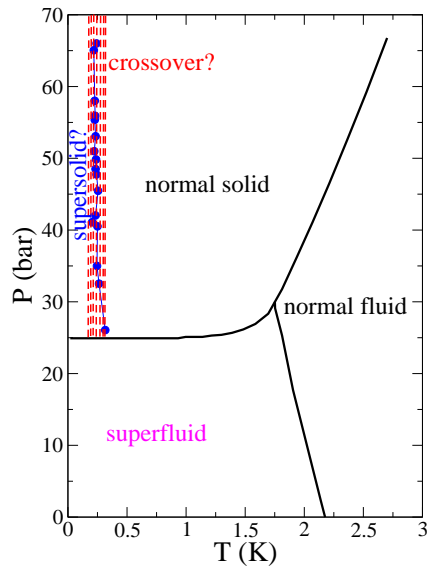}
\end{minipage}
\end{center}
\end{figure}

We provide a  brief  critical analysis of some of the existing data and point out that a significant fraction of the data on mechanical, thermodynamic and  dielectric properties of \HeFour\ can be analyzed in terms of the emergence of  a dissipative {\em viscous} component that  we shall call {\em glassy} component
\cite{Balatsky07,Nussinov07,Graf08,Su10a,Su10b}.
This extra component by itself is sufficient to modify numerous properties of solid \HeFour\ and can be responsible for anomalous thermodynamic, elastic and dielectric
properties in solid \HeFour\   observed in experiments.
The glassy component is not a supersolid in the classical sense \cite{ProkofevAdvances}, yet it can coexist and couple to a supersolid component as some of the proposed superglass phases indicate
\cite{Hunt09,  Boninsegni06,wu2008,Biroli08}.

The precise nature of the state of solid \HeFour\ at lowest temperatures remains a puzzle.  In the past, investigation of \HeFour\ has  played an  important role in  the  development of basic concepts in modern condensed matter physics like superfluidity, order parameter, topological excitations, and critical exponents. Given such a prominent role it played in the past it is paramount to come up with the resolution of the puzzles that are clearly seen in experiments at lowest temperatures. Yet there is one important difference that might be key to a solution this time. We propose that precisely because the compound is clean and very well characterized  the enabling component for the anomalies at lowest temperature are defects that undergo freeze-out and constitute a glass like component.

To sharpen this point of the discussion, any supersolid component would imply some sort of two-fluid hydrodynamics that schematically equates the total mass current
of helium atoms as a sum of normal component and superfluid component determined by the normal and superfluid mass fractions $\rho_n$ and $\rho_s$ and respective velocities. In this two-fluid picture the total mass current is given as
\begin{equation}
{\bf j} = \rho_n {\bf v_n} + \rho_s {\bf v_s}
\end{equation}
In any superfluid or supersolid phase the coefficients $\rho_n(T,P)$ and $\rho_s(T,P)$   are functions of thermodynamic variables like temperature $T$ and pressure $P$.
There is no direct evidence of either dc or ac supersolid mass current at lowest temperatures, where the putative supersolid state  sets in
\cite{Rittner06,Beamish05}, except for the experiments by Hallock \cite{Ray08,Ray09,Ray10,Vekhov2012}. However, it remains to be seen
whether Hallock's results of mass flow and fountain effect cannot be explained through the presence of superleaks connecting the superfluid leads through solid helium with channels.
Therefore, some even proposed that one cannot ``squeeze a superfluid component out of a stone''
 \cite{Dorsey06}.
We know that these supersolid expectations
do not apply in the  case of solid \HeFour\ at lowest temperatures, because TO experiments clearly indicate  that the period and damping exhibit significant hysteresis effects and strong dependence on the history of sample preparations and annealing. All these experimental observations combined would suggest to an impartial observer that there is at the very least, in addition to supersolidity, another physical component at play in \HeFour. Our analysis of various experiments suggests that the crossovers, seen in the specific heat, TO, shear modulus and dielectric function experiments
\cite{Aoki07,LinChan07, Beamish10,Yin11}
are a reflection of the physics whose origin is not due to supersolidity, but a consequence of the dynamics of defects in solid \HeFour\ .

Specifically, we propose the presence of a glassy component in solid \HeFour\ at low temperatures
in order to explain the observed anomalous linear temperature dependence in the specific heat of an otherwise perfect Debye solid
\HeFour\ \cite{Balatsky07, Graf08}.
We used similar ideas when we analyzed the TO, shear modulus,  and dielectric properties  by assuming the presence of a glassy component at the parts-per-million ({\em ppm}) concentration and asked what the dynamic consequences should be.
With the wealth of data available we do not attempt to provide a complete overview of the field, but give a summary of the work centered on the role of a glassy component in an otherwise nearly perfect crystal.

The exact nature  of the glassy component is  not known. For example, it may be caused by
two-level systems of pinned dislocation lines, vortex excitations, etc. It is however important to point out that  the amplitude of period shift can be changed dramatically and depends on growth history and annealing procedures of the crystal. To explain the puzzling features of solid \HeFour \ we \emph{conjecture} that  structural defects like localized dislocation segments or groups of displaced (out-of-equilibrium) atoms effectively form a set of two level systems (TLSs) which are present at low temperatures. These immobile crystalline defects will affect the {\em thermodynamics} of solid \HeFour \cite{Balatsky07,Graf08}, the {\em mechanics} of TOs \cite{Nussinov07} and shear modulus
\cite{Su10a,Su10b},
and dielectric properties \cite{Su11}. Other mechanisms for TLS and glassy dynamics, e.g., due to point defects and grain boundaries are possible as well. We are at a stage where phenomenology allows us to make progress with testable predictions, while a microscopic picture of crystal defects and interactions is still missing.

Early on it was recognized that pinned vibrating dislocation lines can account for a plethora of anelastic damping phenomena in the ultrasound, TO, and shear modulus experiments. Since most experiments are believed to be in the linear or elastic strain-stress regime any plastic deformation due to the motion of gliding dislocation loops has been neglected. However, this is not necessarily so. We proposed \cite{Caizhi2012} that dislocation-induced anomalous softening of solid \HeFour\ is possible due to the classical motion of gliding dislocation lines in slip planes. This picture of dislocation motion is widely accepted in conventional metals. Similar effects are at play in the quantum arena where mobile dislocations (dislocation currents) lead to a screening of applied shear via a Meissner-Higgs type effect
\cite{Zaanen04,Cvetkovic08}. Such unpinned dislocations that screen shear render the system more fluid and may, in line with the framework that we advance in this article,
trigger the \HeFour\ anomalies as the mobile dislocations become quenched at low temperatures \cite{dislocation_core3, Caizhi2012}.  Recent experiments provide further impetus for such a picture
\cite{eyal12a,eyal12b}.

The technique of choice for interrogation of solid \HeFour\ has been the TO  with varying degrees of complexity. However, the TO does not provide any direct information on the microstructure of samples. More direct structural x-ray and neutron measurements do not have the adequate resolution to detect any changes at the {\em ppm} level in the structure of \HeFour\ at lowest temperatures. Additional challenges arise given how small the volume fraction of the glassy component can be. We estimate it to be in the range of
few
hundreds of {\em ppm} in the specific heat and pressure contribution. Therefore the precise characterization of the microstructure of solid \HeFour, growth history, annealing, and \HeThree\ dependence remain pressing issues in resolving the hypothesis of the presence of a glassy component and TLS in  solid \HeFour.
The notion of importance of the role of disorder in solid \HeFour\ received further support over the years in observations of the strong dependence of TO results on the specific design of sample cells, see work  by the group of Chan \cite{DYKim2012}. It is hard to imagine that an intrinsic material property like supersolidity should depend strongly on the stiffness or geometry of the torsional oscillator apparatus, while crystalline disorder can easily be affected by those design properties.

In the analysis of the observed excess specific heat,  we used a model of independent TLS, which gives the canonical signature of a linear in temperature contribution at lowest temperatures. Building on the presence of TLS we  evaluated the mechanical  properties of  the TO using  a model of quenched defects. This approach allows us to make predictions on the viscoelastic properties of \HeFour\ and on the electro-elastic coupling that can be tested in a setup that does not require the TO and hence can be tested over a much broader frequency range. These  predictions allow to directly  verify the very presence of quenched defects in  \HeFour. We therefore would welcome any direct tests of our ideas.

The remainder of this paper is organized as follows. We  present the  general discussion on the role of quenched defects and disorder in solid \HeFour\ with particular attention to the consequence of defects on the dynamics in Section \ref{quenched}. This is followed in Section \ref{thermo} by an analysis of the thermodynamic properties of solid \HeFour. In Section \ref{general_formalism}, we present a unified framework to analyze dynamical response function that invokes arbitrarily high order backaction effects of defects onto the solid bulk. We then invoke, in Section \ref{tos} this approach and  summarize our analysis of the torsional oscillator. In Section \ref{shears} we discuss the shear modulus analysis and the strain-stress relations using a viscoelastic model designed to capture the anelastic contribution from defects. In Section \ref{diels} we discuss predictions for the dielectric properties that follow from the viscoelastic model with locally frozen-in defect dynamics. Finally we conclude with a discussion and give our view on future perspectives in the field.

\section{Defect Quenching and Its Implications}
\label{quenched}

In this review, we will examine the general consequences of a transition
from mobile defects (or dynamic fluid-like components) at high temperatures to quenched immobile defects at low temperatures.
Microscopically, these defects may be dislocations, grain boundaries, vortices, or others.
It may be posited that in quantum solids such as \HeFour\ , zero point motion leads to larger dynamics than is
common in classical solids. To put the discussion in perspective, we recall a few rudiments concerning annealing and quenching.
When quenched, systems fall out of equilibrium.
En route to non-equilibrium states, relaxation times increase until
the dynamic components essentially ``freeze'' (on pertinent experimental time scales) into an amorphous
state. In materials with (sufficiently large) external disorder,
quenching may lead to ``spin-glass'' characteristics. By contrast, in the absence of imposed external disorder, when
fluid components (either classical or quantum) fall out of equilibrium by sufficient rapid cooling (so-called ``super-cooling'')
to low enough temperatures, the resultant state is termed a ``glass''.  As liquids are supercooled, their characteristic relaxation times and viscosity may increase dramatically.
If, instead, the temperature is lowered at a sufficiently slow rate, the system does not quench and instead remains
an ``annealed'' system in thermal equilibrium. Notwithstanding exciting progress, exactly how the dynamic components evolve in  \HeFour\
crystals as the temperature is lowered is, currently, an open problem. An initially surprising and undeniable feature that has, by now, been seen in
many experiments is that the solid  \HeFour\  anomalies depend dramatically on the growth history of the crystal
and diminish as the system is cooled down slowly and defect quenching is thwarted. Memory effects reminiscent to those in
glasses are further present.
These observations imply that ``there is more to life'' than static NCRI and other annealed supersolid properties on their own--
quenching plays a definitive role in triggering the observed effects.
To understand the experimental observations and build a predictive theoretical framework, we invoke
general physical principles
allowing a computation of response functions
and using as input
known characteristics of quenching.
Physically, as alluded to in the Introduction, the quenching is that of the mobile defects (which may constitute only {\it a tiny fraction} of the system) as they become arrested against {\it a crystalline background}.
Our analysis does not rule out the presence of a small supersolid component. The observed large change in the TO dissipation cannot be solely described
by uniform Bose-Einstein condensation \cite{Nussinov07, Graf08, Huse07}.
It remains to be seen if nonuniform Bose-Einstein condensation alone either along grain
boundaries \cite{Clark06} or along the axis of screw dislocations \cite{Boninsegni06, Pollet07, Boninsegni07}
can explain the dynamic response of TOs. Note that simple estimates of the supersolid fraction of dislocation cores are orders of magnitude too small.

There exists a vast literature on defect quenching in systems that range from vortices in superfluid Helium to cosmic strings \cite{Kibble76,Zurek85} and countless others. Our particular focus is, of course, on defects in a crystalline
system (solid \HeFour). Quenched dynamics of such defects is associated with a change of plasticity and related internal dissipation.
Dielectric (and other) response functions in systems of varying plasticity, such as various glass formers as their temperature is lowered,
indicate, in a nearly universal fashion, the presence of a distribution of local relaxation times. These lead to the canonical Cole-Cole or Cole-Davidson distribution functions and related forms as we briefly elaborate.
In an overdamped dissipative system, an impulse (e.g., an external electric field or an elastic deformation) at time $t=0$ leads
a response $g(t)$ which at later times scales as
$g_{single} \sim  \exp(-t/\tau)$ where $\tau$ is the (single) relaxation time. When Fourier transformed to the complex frequency ($\omega$) plane, this leads
to the Debye relaxor $g_{single}(\omega) = g_{0}/(1-i \omega \tau)$.  Now, in systems that exhibit a distribution $P(\tau')$ of local relaxation events, the response functions attain the form
$\int d\tau' P(\tau') \exp(-t/\tau')$. Empirically, in dissipative plastic systems, relaxations scale as $(\exp(-t/\tau)^{c})$ with a power $0<c<1$ that
leads to a ``stretching'' (slower decay) of the response function as compared to its single overdamped mode form of $\exp(-t/\tau)$.
This stretched exponential and other similar forms of the response function capture the quintessence of the distribution of
relaxation times. Two widely used
relaxation time distributions are the Cole-Cole (CC) and Davidson-Cole (DC) functions that describe a
superposition of overdamped oscillators (Debye relaxors) \cite{Phase1, Phase2}.
With $g(\omega) = g_{0} G(\omega)$,
where $g_{0}$ is a material specific constant,
these two forms are given by different choices for the function $G$,
\begin{eqnarray}
\label{CCD}
G_{CC}(\omega)  = 1/[ 1- (i \omega \tau)^{\alpha}], \nonumber
\\ G_{DC}(\omega)  = 1/[ 1- i \omega \tau]^{\beta}.
\end{eqnarray}
Values of $\alpha$ and $\beta$ that differ from unity qualitatively play the role of the real-time stretching exponent $c$.
In the dc limit the mechanical motion of any mobile component will have ceased and there will be no relative motion and no transients. Therefore the coefficient $g_0$ is generally a function of frequency. In the case of the single TO its value is set by the resonance, $g_0 \approx g_0(\omega_0)$.

These relaxation times can be associated
with a distribution of TLSs describing viable low temperature configurations of the defects.
The simple TLS analysis can  account for thermodynamic measurements.
Recent work \cite{Vural11}
obtained results beyond the TLS model with fewer parameters for generic non-uniform systems irrespective of specific microscopic
origin. We will, for the sake of simplicity, review our work on the low temperature properties of \HeFour\ assuming TLSs.
We conjectured  \cite{Balatsky07} that structural defects, e.g.,
localized dislocation segments, form such a set of TLSs at low temperatures.
These immobile crystal defects affect the {\em thermodynamics} \cite{Balatsky07,Graf08}  of bulk \HeFour\
and the {\em mechanics} \cite{Nussinov07} of the TO loaded with \HeFour.
For the analysis of the specific heat, we used independent TLS to obtain the universal signature of a
linear-in-temperature specific heat term at low temperatures.

\section{Thermodynamics}
\label{thermo}

Any true phase transition, including supersolid,  is accompanied by a thermodynamic signature. Therefore it was anticipated that thermodynamic measurements will resolve the existing puzzles of supersolidity. The search for such thermodynamic signatures proved to be challenging  so far,  see e.g.,  measurements of the specific heat
\cite{Swenson62,Frank64,ClarkChan05,LinChan07,LinChan09,WestChan09},
measurements of the pressure dependence of the melting curve \cite{Todoshchenko06,Todoshchenko07},
and pressure-temperature measurements of the solid phase \cite{Grigorev07,Grigorev07b,Lisunov2011}.
The main difficulties lie in measuring small signals at low temperatures in the presence of large backgrounds.
With improving experiments measurements were conducted down to 20 mK.
While there is still no clear evidence of a phase transition in the melting curve experiments, recent pressure measurements and specific heat measurements have both shown deviations from the expected pure Debye lattice behavior.
Early on we proposed that these deviations might be related to a glass transition and be described by the contributions of two-level systems (TLS). \cite{Balatsky07,Graf08}

We model the system of noninteracting TLS with a compact distribution of two-level excitation spacings, see Fig.~\ref{fig:DOS}.
Our results show that the low-temperature deviations in the measured specific heat can be explained by contributions from a
glassy fraction and/or TLS of the solid.


\subsection{Two level system model for the specific heat}

\begin{figure}
\begin{center}
\parbox[t]{0.5\linewidth}{
\caption{Density of states (DOS) of the two-level tunneling system. The black-dashed line represents the DOS for the standard glass model \cite{Anderson72,Phillips72,Balatsky07},
while the blue-solid line is the truncated DOS used to describing the TLS with a cutoff energy.
}\label{fig:DOS}
} \hfill
\begin{minipage}{0.45\linewidth}
\includegraphics[clip, width=1.0\linewidth,angle=0,keepaspectratio]{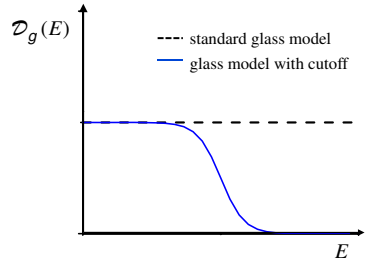}
\end{minipage}
\end{center}
\end{figure}

To avoid complications due to the presence of \HeThree\ atoms, we will compare the effect of different growth processes on ultrapure \HeFour\ containing at most (nominally) 1 ppb of \HeThree\ atoms. At such low levels of impurities, we expect to see the intrinsic properties of solid.

We postulate that at temperature much below the lattice Debye temperature, the specific heat of solid \HeFour\ is  described by
\begin{eqnarray}
C(T) = C_{L}(T) + C_{g}(T) ,
\end{eqnarray}
where the lattice contribution to the molar specific heat is given by $C_L(T) = B_L T^3$,  with coefficient
$B_L = 12 \pi^4 R/5 \Theta_D^3$, $R=8.314$ J/(mol K) is the gas constant, and $\Theta_D$ is the Debye temperature.
The second term describes the glass contribution due to the TLS subsystem  and is given by
$C_{g} (T) = k_B R \frac{d}{d T} \int_0^\infty dE \, {\cal D}_{g}(E) \, f(E)  ,$
with $k_B$  the Boltzmann constant and $f(E)$  the Fermi function.
The density of states (DOS) of the TLS may be modeled by the box distribution function (Fig.~\ref{fig:DOS}):
\begin{eqnarray}
\label{DOE}
{\cal D}_{g} (E) = \frac{1}{2}{\cal D}_0 \left[ 1-\tanh((E-E_c)/W) \right]  .
\end{eqnarray}
Here ${\cal D}_0$ is the zero-energy DOS, $E_c$ is a characteristic cutoff energy,
and $W$ is the width of the truncated density of states. For $E_c \to \infty$, one obtains the standard hallmark result of glasses at low temperatures: $C_{g}(T) = B_g T$,
where $B_g=k_B R {\cal D}_0$. As we will elaborate in the next section, the glass coefficient $B_g$ has an intrinsic finite value at low temperature even for the purest $^4$He samples,
independent of \HeThree\ concentration.

As shown in (\ref{DOE}), our model goes beyond the standard glass model by introducing a cutoff in the DOS of the TLS (Fig.~\ref{fig:DOS}). The cutoff could be due to the finite barrier height of double-well potentials giving rise to the TLS because in real materials the tunneling barrier has an upper bound set by lattice and dislocation configurations \cite{Jaeckle72}. At high temperatures, the TLS contribution is less important since the thermal energy easily overcomes the barrier and effectively resembles a single harmonic degree of freedom.

\begin{figure}
\begin{center}
\parbox[t]{0.30\linewidth}{
\caption{$\delta C/T$ for experiments (squares) \cite{LinChan09} and the modified glass model with a cutoff energy
in the TLS DOS (blue line) of four samples with different structural quality
where $\delta C = C - C_L$.
Note the large deviation of data points at high temperatures in the highest purity crystal SL34 of solid-liquid coexistence,
which makes the subtraction of the Debye contribution in this sample questionable.
}\label{fig:CovT}
} \hfill
\begin{minipage}{0.65\linewidth}
\includegraphics[clip, width=1.0\linewidth,angle=0,keepaspectratio]{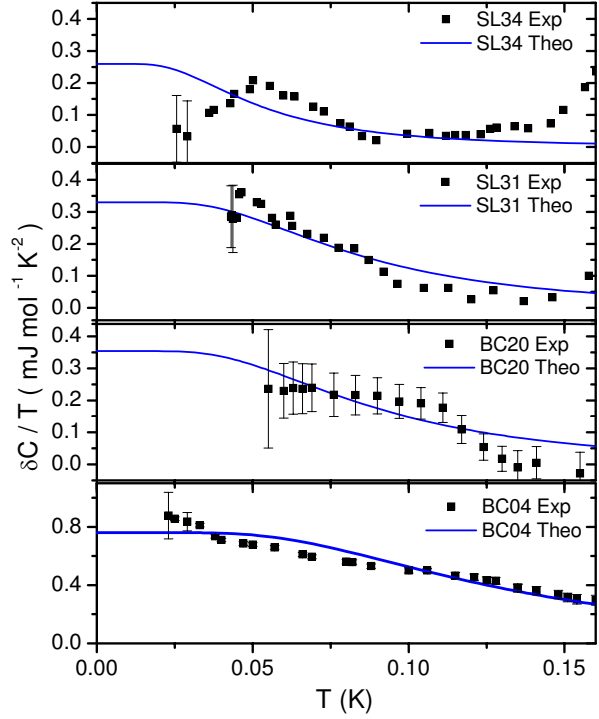}
\end{minipage}
\end{center}
\end{figure}

\subsubsection{Specific Heat}
We compare our calculated specific heat with the experimental data by the Penn State group \cite{LinChan09,LinChan07} for four different
growth processes: BC20, BC04, SL34 and SL31. BC20 (04) is the sample grown by blocked capillary (BC) method over 20 (4) hours. SL 34 (31)
 represents  the samples in solid-liquid coexistence state with 34 (31) percents of solid ratio. Notice that sample SL34 actually corresponds to their reported 75\% solid-liquid coexistence sample and SL31
corresponds to their constant pressure sample (CP)
\cite{Su10a,LinChan09}.
The data are described with three parameters: ${\cal D}_0$, $E_c$ and the Debye temperature $\Theta_D$.
We first determine $\Theta_D$, or the lattice contribution, from the high-temperature data. 
The lattice contribution is then subtracted from $C$ to obtain the difference $\delta C = C - C_L$. We fit $\delta C/T$ with our specific heat formula for TLS.

Next we plot $\delta C/T$ in Fig.~\ref{fig:CovT}.
The TLS model with cutoff describes well the  data. In these plots we fixed the width of the cutoff to $W = 1 \, {\rm \mu eV}$ since there is no qualitative difference when varying $W$ within reasonable range.
Notice that the shape of $\delta C(T)$ depends strongly on the subtraction of the high-temperature lattice contribution.

The TLS behavior is mainly characterized by the zero-energy DOS and the cutoff energy of the TLS, which are both noticeably larger in BC04,
see Table~\ref{para_table}. This may be explained by the rapid growth process of a strained crystal, which gives rise to both a larger TLS concentration, $n_{\rm TLS}$, and a smaller cutoff energy, $E_c$, i.e., a smaller maximum tunneling barrier height. Since the TLS concentrations of these samples range from $3.7$ to $21.5$ {\em ppm}, which are at least 1000 times larger than the nominal $^3$He concentration, we believe that $^3$He has
no effect on the observed intrinsic heat capacity of ultrapure solid $^4$He.

\begin{table}[h!]
\begin{center}
\begin{tabular}{c|ccccccc}
& $P$ & $V_m$ & $\Theta_D $ & ${\cal D}_0 \times 10^4$ & $E_c \times 10^2$
&$n_{\rm TLS}$ & $\Delta S$\\
& (bar) &(cm$^3$/ mol)& (K) & (1/meV) & (meV) & ({\em ppm}) & ($\mu$J/(mol K)) \\
\hline
SL34      & 25 & 21.25 & 24.5 & 2.2 & 1.7 & 3.7 & 21.3\\
SL31      & 25 & 21.25 & 24.8 & 2.9 & 2.2 & 6.4 & 36.9\\
BC20      & 33 & 20.46 & 29.7 & 3.0 & 2.3 & 6.9 & 39.5\\
BC04      & 33 & 20.46 & 28.9 & 6.5 & 3.3 & 21.5& 115.0\\
\end{tabular}
\end{center}
\caption{Physical and model parameters: Debye temperature $\Theta_D$, zero energy TLS DOS
${\cal D}_0$, cutoff energy $E_c$, concentration of TLS $n_{\rm TLS}={\cal D}_0\times E_c$ and excess entropy $\Delta S$. }
\end{table} \label{para_table}

\subsubsection{Entropy Analysis}
Our analysis of the excess entropy supports the existence of a glassy component or TLS. The excess entropy,
\begin{eqnarray}
\Delta S(T) = \int_0^T dT' \,\delta C(T')/T',
\end{eqnarray}
is associated with an excess specific heat.
We find consistently for specific heat measurements \cite{ClarkChan05,LinChan07,LinChan09} that the
obtained entropy values
$\Delta S \sim 10^{-4}$
J/(K mol) are 5 to 6 orders of magnitude smaller compared to the theoretical prediction for a homogeneous supersolid
if the entire sample actually underwent Bose-Einstein condensation (BEC).  In the limit of a non-interacting BEC
 bosons with a quadratic dispersion one finds the standard result
$\Delta S_{BEC}=15/4 (\zeta(5/2)/\zeta(3/2))\, R (T/T_c)^{3/2}\sim (5/4)\, R\sim 10.4$ J/(K mol).
This means that if $\Delta S$ is indeed due to supersolidity, then the supersolid volume fraction is at most
11 {\em ppm} or 0.0011\% in the most disordered or quenched sample of the four ultrapure samples studied in this work,
i.e., sample BC04.
Such a supersolid fraction in the specific heat is more than 100 to 1000 times smaller than is usually reported for the
non-classical rotational inertia fraction (NCRIF) in TO experiments.
This enormous discrepancy between supersolid fractions in specific heat and TO experiments was already
noticed in Refs.~\cite{Balatsky07,Graf08}, while Lin et al.\ \cite{LinChan09} keep invoking a hyperscaling mechanism
of unknown origin.
Until to date, this discrepancy remains a major puzzle that is hard to reconcile within a supersolid scenario.

The validity of the analysis of the entropy in terms of a non-interacting BEC is repeatedly questioned on grounds of how robust it is in the presence of interactions. We discussed in our original study \cite{Balatsky07} the effects of interactions on the entropy. Here it is
important to realize that the entropy is
a total count of all low-energy states irrespective of a particular model for the specific heat. Hence, we concluded that strong-coupling effects
cannot change the order of the effect. They may only change the magnitude. For example,  the
well-known strongly correlated superfluid system \HeFour\ possesses a superfluid entropy of $\sim 0.6 R \sim 4.6 $ J/(K mol) \cite{Ahlers1973},
which is only half the value of the non-interacting BEC. With hindsight this justifies the neglect of strong-coupling effects in the order of magnitude analysis.
Clearly the entropy is a reliable measure of any phase transition in  \HeFour. No matter how one looks at this puzzle, the reported excess entropy is either far too small to explain observed NCRIF effects in torsional oscillators or  far too large to describe the boiling off of \HeThree\ atoms from dislocation lines, when the nominal concentration of \HeThree\ is less than 1 {\em ppb}. For those reasons, we argued in favor of two-level systems of low-lying states until a better microscopic understanding of solid \HeFour\ emerges.

\subsubsection{Comparison with Pressure Measurements}

\begin{figure}
\begin{center}
\parbox[t]{0.35\linewidth}{
\caption{($P-P_0)/T^2$ vs. $T^2$ for $P \sim 33$ bar. The squares represent the data reported by Grigor'ev et al. \cite{Grigorev07,Grigorev07b} The blue-thick and black-thin lines are predictions for BC04 and BC20, respectively.
The black-dotted line has been shifted vertically by a constant compared to BC04 to illustrate the capability of describing
Grigor'ev's data by the TLS model. Predicted curves for BC04 and BC20 are shown for comparison. }\label{fig:PovTsq2}
} \hfill
\begin{minipage}{0.64\linewidth}
\includegraphics[width=1.0\linewidth,angle=0,keepaspectratio]{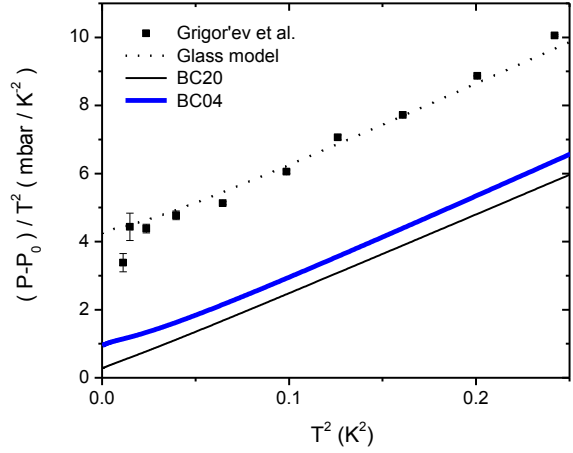}
\end{minipage}
\end{center}
\end{figure}

\begin{figure}
\begin{center}
\parbox[t]{0.35\linewidth}{
\caption{Low-temperature pressure deviation from lattice contribution of Debye solid (data by Yin et al.\ \cite{Yin11}). The intercept of $\Delta P/T^2$ vs.\ $T^2$ extrapolated from low-temperature data points is in agreement with the TLS contribution of order 100 {\em ppm} ($\Delta P = P-P_0$).
The arrow marks the deviation from the Debye solid behavior. The large scatter in data  at lowest $T$ is due to the subtraction of  $P_0$. Dashed lines are guides to the eyes.
}\label{fig:PT}
} \hfill
\begin{minipage}{0.64\linewidth}
\includegraphics[width=1.0\linewidth,angle=0,keepaspectratio]{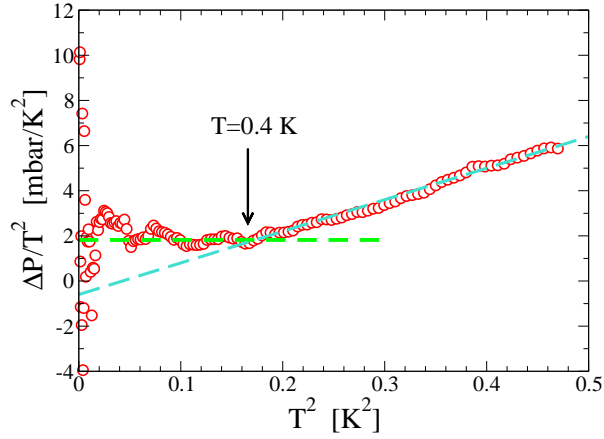}
\end{minipage}
\end{center}
\end{figure}

Next we relate the pressure measurements with the specific heat measurement through our model.
The quantities to characterize the pressure measurement in the combined lattice and glass models are $a_L$ and $a_g$ defined by
\begin{eqnarray}
P(T) \equiv  P_0+P_L(T)+P_g(T) = P_0 + a_L T^4+ a_g T^2  ,
\end{eqnarray}
where $P(T)$ is the pressure at temperature $T$.
$P_0$, $P_L$, $P_g$ are the corresponding pressure contributions of the ions at zero temperature, lattice vibrations, and two-level  excitations of the glass. On the other hand, the thermodynamic Maxwell relations between pressure and specific heat give
\begin{eqnarray}\label{maxwell}
\left(\frac{\partial P}{\partial T} \right)_V =
\frac{\gamma_g}{V_m} \, C_{g,V} + \frac{\gamma_L}{V_m} \, C_{L,V} ,
\end{eqnarray}
where $\gamma_i$ are the Gr\"uneisen coefficients of the glass  ($g$) and lattice ($L$).
Literature values for the Gr\"uneisen coefficient of phonons in solid hcp $^4$He range between $2.6 < \gamma_L < 3.0$
\cite{Grigorev07b,Driessen86}, while nothing is known about $\gamma_g$ of glassy $^4$He.
For simplicity we assume $\gamma_g \sim \gamma_L =2.6$.
Equation~(\ref{maxwell}) is related to the first Ehrenfest relation involving the compressibility, which was shown to be always satisfied in glasses \cite{Nieuwenhuizen1997}.
In Figs.~\ref{fig:PovTsq2} and \ref{fig:PT} we show the temperature dependence of the $(P-P_0)/T^2$ data reported by Grigor'ev et al.
\cite{Grigorev07,Grigorev07b} and Yin et al.\cite{Yin11}.
The curves for samples BC04 and BC20 are derived from our specific heat analysis of the data by Lin et al.\cite{LinChan09}
The key result is the  dependence $\sim T^2$ with finite intercept at $T=0$.
In the TLS model, the finite intercept describes the glassy contribution, whereas the $T^2$ behavior is attributed
to phonons.
In conclusion, the data by Grigor'ev et al.\  and Yin et al.\ are in agreement with predicted $(P-P_0)/T^2$
curves for a system of TLS.

\section{General response function formalism and physical origin of dynamical anomalies}
\label{general_formalism}

To rationalize the TO experiments, we developed a general phenomenological formalism in Ref.~\cite{Nussinov07}. We have since extended and applied it to visco-elastic and dielectric properties. With simple modifications, this predictive approach can be used to study all measurable dynamical response functions. Here, we summarize the essence of our approach. In later sections, we will apply it in a self-contained way to the various
quantities that we wish to interrogate.  We start with the equation of motion for generalized coordinates $q_{i}$. These coordinates may be an angle, $q=\theta$,  in the case of the TO experiments, Cartesian components of atomic displacements , $q_{i}=u_{i}$ with ${\bf u}$ the local atomic displacement (as in our analysis of the visco-elastic and dielectric response functions), or any other. Associated with these coordinates are conjugate generalized momenta $p_{i}$ (e.g., angular momentum in the TO analysis, linear momentum for Cartesian coordinates) and their associated generalized forces $F_{i}$ (torques in the case of the TO, and rectilinear forces for atomic displacements). Physically, these forces correspond to a sum
of two contributions:
 (i) Direct forces ($F_{direct}$). These may originate from either externally applied forces on the solid ($F_{ext}$) or lowest order ``direct'' internal forces $(F_{int}$). By direct forces we allude
 to forces on the coordinate $q$ that do {\it not} involve the response of the system on $q$ as a result of its change.

 (ii) Indirect ``backaction'' forces ($F_{BA}$). These allude to higher order effects wherein a variation in the coordinate $q$ can lead to displacements in other parts of the medium (e.g., those involving plastic regions or
nearby atoms) which then act back on the original coordinate $q$.
To linear order
\begin{eqnarray}
\label{BAE}
F_{BA}(t) = \int_{-\infty}^t g(t-t') q(t') dt'.
\end{eqnarray}
The backaction function $g(t-t')$ captures how a displacement $q$ at time $t'$ can lead to a perturbation in the solid
which then acts back on the coordinate $q$ at time $t$.  With these preliminaries in tow,
we now outline our standard linear response formalism which further takes into account all higher order backaction effects.

$\bullet$ (1) Write down the Newtonian equation(s) of motion for the generalized coordinate(s) $q$ that we wish to study.
With $\chi_{0}$ a suitable differential operator $\chi_{0}^{-1}$, these can be cast as
\begin{eqnarray}
\chi_{0}^{-1} q(t) =  F_{direct}(t) + F_{BA}(t).
\label{general_EOM}
\end{eqnarray}
This equation might seem a bit formal. To make it concrete, we note that in the simplest case that we will discuss- that of the compact scalar angular coordinate $\theta$ for the TO orientation,
the operator
\begin{eqnarray}
\chi_{0}^{-1} = I_{osc} \frac{d^{2}}{dt^{2}} + \gamma_{osc} \frac{d}{dt} + \alpha_{osc}
\end{eqnarray}
where $I_{osc}, \gamma_{osc},$ and $\alpha_{osc}$ are, respectively, the oscillator moment of inertia,
dissipation, and
torsion rod
stiffness.  A more complicated tensorial operator involving the elastic modulus
appears when writing the equations of motion for the Cartesian components of the atomic displacements.

$\bullet$ (2) When Fourier transformed to the complex frequency ($\omega$) plane, $\chi_{0}(\omega)$ corresponds to
the bare susceptibility. We will denote the Fourier transforms of the various quantities (forces, backaction function) by simply making it clear that
the argument of the various quantities is now the frequency $\omega$ and not the time $t$. We trivially recast Eq. (\ref{general_EOM}) as
\begin{eqnarray}
\chi^{-1}(\omega) q(\omega) = F_{direct}(\omega),
\end{eqnarray}
where
\begin{eqnarray}
\label{Dyson}
\chi^{-1}(\omega) = \chi_{0}^{-1}(\omega) - g(\omega).
\end{eqnarray}
Equation (\ref{Dyson}) will be used in all of our upcoming analysis.
The physical content of Eq. (\ref{Dyson}) can be seen by writing
its inverse
as a geometric (or Dyson) series
\begin{eqnarray}
\label{longc}
\chi(t) = \chi_{0}(t) + \int dt' \chi_{0}(t) g(t-t') \chi_{0}(t') \nonumber
\\ + \int dt' \int dt" \chi_{0}(t) g(t-t') \chi_{0}(t') g(t'-t") \chi_{0}(t") + ... .
\end{eqnarray}
The terms in this series correspond to
(a) the direct contribution  ($\chi_{0}$),
(b) a lowest order backaction effect wherein a displacement at an earlier time $t'<t$ leads to a deformation of the solid which then acts back on the coordinate at time $t$ (the term $\int dt' \chi_{0}(t) g(t-t') \chi_{0}(t') $),
(c) a higher order process in which a deformation at time $t''$ leads to a backaction from the solid on the coordinate at time $t'>t"$ which in turn then acts back on its surroundings which then acts back on the original coordinate at time $t$ (the term $\int dt' \int dt" \chi_{0}(t) g(t-t') \chi_{0}(t') g(t'-t") \chi_{0}(t") $), and so on ad infinitum.
In Fourier space, the convolution integrals become products and Eq. (\ref{longc})
becomes
\begin{eqnarray}
\label{dyson}
\chi(\omega) = \chi_{0}(\omega) + \chi_{0}(\omega) g(\omega) \chi_{0}(\omega) + \chi_{0}(\omega) g (\omega) \chi_{0}(\omega) g(\omega) \chi_{0}(\omega) + ... .
\end{eqnarray}
The geometric series of Eq. (\ref{dyson}) sums to Eq. (\ref{Dyson}).
As simple as it is,
Eq. (\ref{Dyson}) combining standard linear response theory (step 1) with backaction effects (or the Dyson equation) of step 2,
leads to a very powerful tool that allows us to investigate numerous systems while accounting for arbitrarily high order backaction effects.
As is well known and we will expand on and employ in later sections, the real and imaginary parts of the poles of the susceptibility $\chi(\omega)$ allow us to
probe typical oscillation times and dissipation. This will allow us to connect with TO and other measurements and make precise statements about
the backaction function $g$ which affords information about the dynamics within the solid. It is important to stress that Eq. (\ref{Dyson}) is general.
In the derivation above no assumptions need to be made concerning the precise physical origin of the backaction function $g$.
The adduced function $g$ captures all effects not present in the direct equations of motion for the normal solid. If supersolid
effects would be present, they will directly appear in the function $g$.

Thus far, our sole assumption was that the deformations $q$ are small enough to justify the linear (in $q$) order analysis of
Equations (\ref{BAE}, \ref{general_EOM}) for measurements on the rigid solid. We now invoke additional assumptions (which have been partially vindicated in a growing number of experiments since our original proposal
\cite{Nussinov07}). These assumption relate to the form of $g(\omega)$ and its dependence on temperature. They are motivated by our view of defect quenching and characteristic relaxation times as the origin of the
\HeFour\ anomalies.

$\bullet$ (3) As we will elaborate on in later sections, data for disparate susceptibilites $\chi$ taken at different frequencies $\omega$ or temperatures $T${\it collapse onto one curve}.
This indicates that $g$ is a function of only one dimensionless argument ($\omega \tau(T)$) instead of both $\omega$ and $T$ independently.
That is, there is only one dominant temperature dependent relaxation time scale $\tau(T)$ for the backaction of the quenched solid on the original coordinate $q$.
As is well known and alluded to in Section (\ref{quenched}), exponential damping with a single relaxation time $\tau$,
leads to a function $g_{single}(\omega)= g_{0} (1- i \omega \tau)/(1+ \omega^{2} \tau^{2})$ which, when plotted with the real and imaginary parts of
$g$ along the horizontal and vertical axes describes a semi-circle as a function of the dimensionless quantity $(\omega \tau)>0 $. However, when plotted
in this way, the \HeFour\ data for the complex susceptibility measured by TO and other probes lead to a collapsed curve which is more like that of a skewed semi-circle
 and there is a distribution of
relaxation times about a characteristic time scale $\tau$. For ease of analysis, we will approximate the complex response
functions by Eqs. (\ref{CCD}).
The curve collapse allows for information about how the characteristic transient relaxation times $\tau(T)$ increase
as $T$ is decreased. We further invoke
the Vogel-Fulcher-Tammann form for glasses \cite{Rault00},
\begin{eqnarray}
\label{VFT_eq}
\tau(T) =
\left\{
\begin{array}{ll}
\tau_0 \,e^{\Delta/(T-T_0)} & \mbox{ for $T>T_0$} , \\
 \infty  & \mbox{ for $T \le T_0$} .
\end{array}
\right.
\end{eqnarray}
Here, $T_{0}$ is the temperature at which the relaxation times would
truly diverge and $\Delta$ is an energy scale.  In fitting the data in this way, negative values $T_{0}<0$ were often found.
That is, the typical dynamics as adduced from the collapsed $\tau(T)$ is faster than
simple activated dynamics (one in which $T_{0}=0$ in Eq. (\ref{VFT_eq})).
This is consistent with the intrinsic quantum character of the solid \HeFour\ crystal
with large zero point motion as compared to classical activated dynamics.

What is the physical content of this general formalism vis a vis the putative supersolid transtition? Given perturbations of a typical frequency $\omega$, the backaction response $g$ from the plastic components acting on $q$
may either be sufficiently rapid or slow to respond. Just at the tipping point when $\omega \tau(T_{X}) \simeq 1$
different components will be maximally out of synchrony with each other in being able to respond to
the perturbation or not and the dissipation (given by the reciprocal of the imaginary part of the zero of $\chi^{-1}(\omega)$) is maximal.
Similarly, there will be a change in the typical periods of the system between a system which at high $T$  (i.e., $T< T_{X}$) contains rapidly
equilibrating plastic components to those at low
$T$, which
are too slow to respond and thus
the system appears to have undergone ``a transition''.

In the sections that follow, we will apply and replicate anew and at great length the considerations outlined above to the particular set of physical quantities that we wish to investigate.
We start, in Section \ref{tos} with the investigation of the TO (for which the above formalism was first developed) and then move to explore other arenas- the viscoelastic (Section \ref{shears})
 and dielectric response (Section \ref{diels}) functions
where the above formalism leads to experimentally testable predictions.

\section{General susceptibility and response function of torsional oscillators}
\label{tos}

A formulation of the rotational susceptibility of the TO was given in Ref.~\cite{Nussinov07}. It is now often used as a basis
for
the linear response discussion \cite{Dorsey08,Pratt11,Gadagkar2012}.
In Section (\ref{general_formalism}) we outlined the key points of this formulation when written in its general form.
Since its derivation it has been applied by us and others to study other response functions.
The result of Eq. (\ref{central}) below is none other than Eq. (\ref{Dyson}) when the generalized coordinate $q$ corresponds to the TO angle. The bulk of this section will be devoted
to analyzing the experimental consequences of this relation and its related counterpart for the double TO.

It is important to realize that the TO experiment measures the period and dissipation of the entire apparatus by reporting the relationship between the force and displacement (angle). Therefore  a model is needed to determine the relation between observable period and dissipation and the moment of inertia, damping and effective stiffness of the media.
We start with the general equation of motion
for a harmonic TO defined by an angular coordinate
$\theta$ in the presence of an external and internal torque,\cite{Nussinov07}
\begin{eqnarray}
I_{osc} \ddot{\theta}(t) + \gamma_{osc} \dot{\theta}(t) + \alpha_{osc} \theta(t)
= {M}_{ext}(t) + {M}_{int}(t).
\label{de}
\end{eqnarray}
Here, $I_{osc}$ is the moment of inertia of the (empty) TO apparatus,
$\alpha_{osc}$ is the restoring constant of the torsion rod, and $\gamma_{osc}$ is its
damping coefficient. $M_{ext}(t)$ is the externally imposed torque by the drive.
${M}_{int}(t) = \int {g}(t-t') \theta(t') dt'$
is the internal torque exerted by solid \HeFour\ on the oscillator for a system with time
translation invariance.  In general, the backaction $g(t-t')$ is temperature, $T$, dependent.

The external torque, $M_{ext}(t) = \dot{L}(t)$, is the derivative
of the total angular momentum of a rigid body,
$L(t) = \frac{d}{dt} \int d^{3}x ~\rho(\vec{x}) r^{2} ~ \dot{\theta}(\vec{x})$,
where $r$ is the distance to the axis of rotation, $\rho(\vec{x})$
is the mass density and $\dot{\theta}(\vec{x})$ the local angular
velocity about the axis of rotation.\cite{Anderson08}
The experimentally measured quantity is the angular motion of the TO -
not that of bulk helium, which is enclosed in it. Ab
initio, we cannot assume that the medium moves as one rigid body.
If the non-solid subsystem ``freezes" into a glass, the medium will move with greater
uniformity and speed. This leads to an effect similar to that of the nonclassical
rotational moment of inertia, although its physical origin is completely different.
We argue for an alternate physical picture, namely that of softening of the oscillator's stiffness.

The angular coordinate $\theta(t)$ of the oscillator is a convolution of the applied
external torque ${M}_{ext}(t)$ with the response function ${\chi}(t,t')$.
Causality demands ${\chi}(t,t') = \theta(t-t') {\chi}(t,t')$. Under Fourier
transformation, this leads to the Kramers-Kronig relations.
In any time translationally invariant system, the Fourier
amplitude of the angular response of the TO is
\begin{eqnarray}
\chi_0^{-1}(\omega)\theta(\omega) =  M_{ext}(\omega) + M_{int}(\omega) .
\label{ft}
\end{eqnarray}
Defining the total angular susceptibility as
$\chi^{-1} = \chi_0^{-1} - M_{int}$,
we write the effective inverse susceptibility as
\begin{eqnarray}
\chi^{-1}(\omega) =
\alpha_{osc} - i \gamma_{osc} \omega - I_{osc} \omega^{2} - g(\omega),
\label{central}
\end{eqnarray}
where $g(\omega)$ is the Fourier transform of the backaction
due to the added solid \HeFour.
In what follows, we will treat the
backaction as a small perturbation to the TO chassis.

We will now apply our formalism to the study of the
ingle TO, which is described by Eq.~(\ref{central}),
and then turn to the double  TO.
Very recently, Beamish \cite{Beamish2012} and Maris \cite{Maris2012} employed the same general linear response formalism
to explain some of the TO results in terms of purely mechanical effects due to either the changing
stiffness of the torsion rod or floppiness
of the sample cell flange (lid).

\subsection{A model for the single torsional oscillator}

In what follows, we analyze the experimental consequences of Eq. (\ref{central}).

\subsubsection{Rotational susceptibility - period and dissipation}
We can now calculate specific consequences of the phenomenological model introduced above.
The effective oscillator parameters are defined as
the sum of parameters describing the apparatus, $\chi_0^{-1}$,
and the added solid \HeFour\  given by
\begin{equation}
g(\omega)=i\gamma_{He}\omega + I_{He}\omega^2 + g_0 G(\omega) .
\label{geq}
\end{equation}
It is convenient to introduce a net moment of inertia $I = I_{osc} + I_{He}$
and net dissipation $\gamma=\gamma_{osc}+\gamma_{He}$.
The transient dynamic response function $G(\omega)$ can be approximated by a distribution of
overdamped oscillators with relaxation time $\tau$ as discussed in Section (\ref{quenched}) [see Eqs. (\ref{CCD}), in particular].

The resonant frequency of the TO with a backaction
is given by the root of
\begin{eqnarray}
\chi^{-1}(\omega) =
\alpha - i \gamma \omega - I \omega^{2}- g_{0} G(\omega) \equiv 0.
\label{central_glass}
\end{eqnarray}
As discussed in Section \ref{general_formalism}, we anticipate that when the relaxation time
is similar to the period of the
underdamped oscillator, the dissipation will be maximal, sometimes referred to as ``$\omega\tau=1$'' physics.

In linear response theory, the homogeneous Eq.~(\ref{central_glass}) is scale invariant.
Thus, we normalize all oscillator quantities by the effective moment of inertia $I$, i.e.,
$\bar{\alpha} = \alpha/I$, $\bar{\gamma} = \gamma/I$, and $\bar{g}_0 = g_0/I$.
As can be seen from Eq.~(\ref{central_glass}), for an ideal dissipationless oscillator ($\bar\gamma=0$),
the resonant frequency
$\omega_{0}= \sqrt{\bar\alpha}$
is the pole of $\chi(\omega)$ in the limit $1/\tau \to 0$. If we expand $\chi^{-1}$ about this root,
$\omega= \omega_{0} + \delta\omega$,
we find to leading order in $\delta\omega$
\begin{eqnarray}
\delta\omega \approx
-\frac{i \bar\gamma \omega_{0} +
\bar{g}_0 G(\omega_0)
}{i \bar\gamma + 2\omega_{0}} .
\end{eqnarray}
It follows that the shift in dissipation with respect to high temperatures is
\begin{eqnarray}
\Delta Q^{-1} \equiv Q^{-1} - Q^{-1}_0
\approx \frac{\bar{g}_0}{\omega_{0}^2} {\rm Im\ } G(\omega_0) ,
\label{Q-1p}
\end{eqnarray}
whereas the shift in resonant frequency is
\begin{eqnarray}
\Delta\omega\equiv 2\pi (f_0-f)
&\approx& \frac{\bar{g}_0}{4 \omega_{0}^2}
\Big(
  2 \omega_{0} \, {\rm Re\ }G(\omega_0) + \bar\gamma \, {\rm Im\ }G(\omega_0)
\Big),
\label{P-1p}
\end{eqnarray}
which increases monotonically when $T$ is lowered.
Combining Eqns.~(\ref{Q-1p}) and (\ref{P-1p}) for the strongly underdamped oscillator,
we arrive at
\begin{equation}\label{Q-P-ratio}
\frac{\Delta Q^{-1}}{\Delta \omega} =
\frac{ 4 {\rm Im\ }G(\omega_0) }{ 2\omega_0{\rm Re\ }G(\omega_0) + \bar{\gamma} {\rm Im\ }G(\omega_0) } \approx
\frac{2}{\omega_{0}} \frac{ {\rm Im\ }G(\omega_0) }{ {\rm Re\ }G(\omega_0) }.
\end{equation}
It is this general relationship for the response function of the damped oscillator that was successfully applied in the
TO analysis by Pratt et al.
\cite{Pratt11,Gadagkar2012}
to demonstrate the interplay of rotational, relaxation,
and shear dynamics in solid \HeFour.
For a Debye relaxor the ratio of Eq.~(\ref{Q-P-ratio}) reduces to $2\tau$ and provides a direct means to measure the
relaxation time.
Similar results for the ratio were obtained for other phenomenological models.\cite{Huse07, Dorsey08}
For example, Huse and Khandker \cite{Huse07} assumed a simple phenomenological
two-fluid model, where the supersolid is dissipatively
coupled to a normal solid resulting in a ratio of
${\Delta Q^{-1}}\frac{\omega_0}{\Delta \omega} \approx 1$, while Yoo and Dorsey\cite{Dorsey08}
developed a viscoelastic model, and Korshunov\cite{Korshunov09} derived a two-level system glass model
that captures the results of the general model originally proposed by Nussinov et al.\cite{Nussinov07}
To make further progress we assume that $\tau$ follows the phenomenological Vogel-Fulcher-Tammann (VFT) equation
of Eq. (\ref{VFT_eq}).

\begin{figure}
\begin{center}
\includegraphics[width=0.60\linewidth,angle=0]{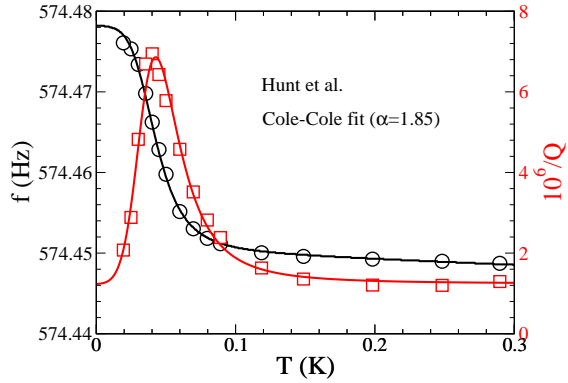}
\end{center}
\caption{The resonant frequency (black, left axis) and dissipation (red, right axis) vs.\ temperature.
The experimental data
from Hunt et al.\cite{Hunt09} are well described by a Cole-Cole (CC) distribution function.
\cite{Graf10,Graf11}
}\label{fig_Hunt}
\end{figure}

Figure~\ref{fig_Hunt} provides a fit to the measured data by Hunt et al.\cite{Hunt09} assuming a CC distribution of
relaxation times. As shown, an excellent
fit is obtained. For comparison, we also tried a DC distribution for relaxation times, but found only fair agreement.
It is worth mentioning that unlike in the Debye relaxation analysis by Hunt and coworkers, i.e., a single overdamped mode,
we do not require a supersolid component to {\em simultaneously} account for frequency shift and dissipation peak
\cite{Graf10,Graf11}.
Our model leads to a universal scaling of period change vs.\ dissipation in a Cole-Cole or Davidson-Cole plot as seen in
Ref.~\cite{Pratt11,Gadagkar2012}.
Indeed similar viscoelastic behavior may have already been observed in solid hydrogen \cite{Clark2006}.

\subsection{A model for the double oscillator}

\begin{figure}
\bigskip
\begin{center}
\includegraphics[ width=0.25\linewidth, keepaspectratio, angle=0 ]{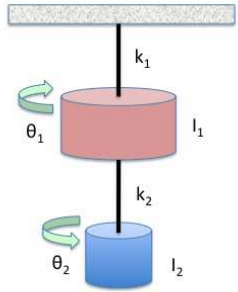}
\end{center}
\caption{Cartoon of the double torsional oscillator modeled in Eqns.~(\ref{eqns}).
The upper moment of inertia ($I_1$) is the dummy bob, while the lower moment of inertia ($I_2$)
is the cylindrical pressure cell that can be loaded with \HeFour. The stiffness of the torsion rods is given by
$k_1$ and $k_2$ with $k_1 \approx k_2$ by design.}
\label{Fig2}
\end{figure}

The double TO results of the Rutgers group by Kojima have proved difficult to explain when
simply extrapolating from the single TO model
\cite{Dorsey08}.
Here we model the coupled double TO, sketched in Fig.~\ref{Fig2},
by the following system of equations:
\begin{eqnarray}\label{eqns}
  \left( -I_1 \omega^2 - i \gamma_1 \omega + k_1 + k_2 \right) \Theta_1(\omega) - k_2 \Theta_2(\omega) &=& F(\omega) ,
\nonumber\\
  \left( -I_2 \omega^2 - i \gamma_2 \omega + k_2 - g(\omega) \right) \Theta_2(\omega) - k_2 \Theta_1(\omega) &=& 0 .
\end{eqnarray}
where $\Theta_i$ are torsion angles, $\gamma_i$ are damping coefficients, $k_i$ are torsion rod
stiffnesses, $g(\omega)$ is the glass backaction term, and $F(\omega)$ is the applied external torque.
The subindex ``$1$'' refers to the upper or dummy bob in the experiment,
while ``$2$'' refers to the lower oscillator with the pressure cell that can be loaded with solid \HeFour.
For a strongly underdamped oscillator and a small backaction, it suffices to solve first for the
bare resonant frequencies and later include perturbatively damping and backaction terms, for details see
Ref.~\cite{Graf10}.
More recently this approach has been extended to a triple TO \cite{Mi2011}.

\begin{figure}
\begin{center}
\includegraphics[ width=0.8\linewidth, keepaspectratio, angle=0 ]{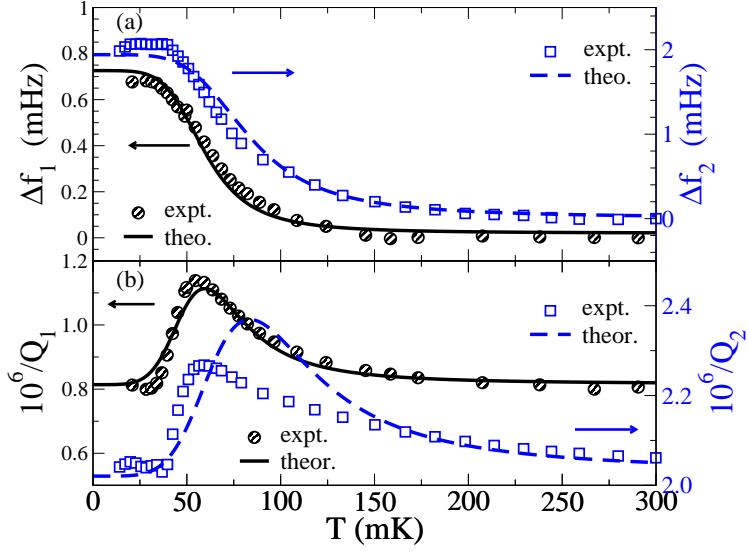}
\end{center}
\caption{Frequency and dissipation in double TO by Aoki et al.\cite{Aoki07}
(symbols) compared with glass theory (lines).
Panel (a): Temperature dependence of resonant frequency shifts $\Delta f_1$ (black, left axis) and $\Delta f_2$ (blue, right axis).
Panel (b): Temperature dependence of dissipation $Q_1^{-1}$ (black, left axis) and $Q_2^{-1}$ (blue, right axis).
The experimental data were corrected for the significant temperature-dependent background of the empty cell.
}
\label{fig_DO}
\end{figure}

Figure~\ref{fig_DO} shows good agreement between our phenomenological model of the coupled double TO and experiment.
The TO parameters $I_i$ and $k_i$ can be determined from the bare resonant frequencies $f_i^0 = \omega_i^0/2\pi$.
In addition, the damping coefficients $\gamma_i$ can be extracted from the high-temperature dissipation $Q_{i, \infty}^{-1}$.
Finally, the backaction
${g}(\omega)$ accounts through $\tau(T)$
for the temperature dependence of
$\Delta f_i$ and $Q_i^{-1}$.
Our phenomenological theory of the double oscillator explained for the first time both
frequency shift and dissipation peak for in-phase and out-of-phase
torsional response \cite{Aoki07}.
Data for in-phase frequency $f_1=496$ Hz and out-of-phase $f_2=1173$ Hz
are shown in Fig.~\ref{fig_DO}, plotted against the
temperature.
The obtained values for moment of inertia and rod stiffness agree well with
other estimates \cite{Aoki08}.

Remarkably an anomalous damping coefficient $\gamma_1 \sim -\gamma_2$
is required to explain the behavior of increased
dissipation with increased frequency. Such anomalous damping
is already required to describe the unloaded pressure cell. Thus it is {\em unrelated} to the
properties of solid \HeFour\ and an intrinsic property of the composite TO.
After loading the cell with solid \HeFour\ the dissipation ratio becomes
$Q_{2}^{-1}/Q_{1}^{-1} = 2.5$ at 300 mK with frequency ratio $f_2/f_1 = 2.37$.
Our fit results in a negative parameter $T_0=-32.73$ mK. This effective negative value of $T_{0}$ is in line with
earlier comment in Section \ref{general_formalism},
concerning the quantum character of solid  \HeFour. This value may be indicative of strong zero point quantum fluctuations that thwart a
glass transition.

Finally, the comparison in Fig.~\ref{fig_DO} shows that an explicit
frequency-dependent backaction  must be used with
$g_0(\omega) = g_0 \left({\omega}/{\omega_1^0}\right)^p$
and $p = 1.77$ to account for the experimental fact of $\Delta f_1/f_1 \approx \Delta f_2/f_2$,
i.e., the relative frequency shift is unaffected by the changing resonant frequency.
In contrast, various theories describing solid \HeFour\ in torsional oscillators as viscoelastic material \cite{Dorsey08}
or two-level systems moving through a solid matrix \cite{Andreev07, Andreev09, Korshunov09}
predict an exponent of $p=4$ for the backaction term.


\section{Shear and stiffness of a viscoelastic medium}
\label{shears}

Another aspect of the dynamic response  of $^4$He crystals was revealed through a series of elasticity studies.\cite{Paalanen81,Goodkind02,Burns93,Beamish07,Beamish09,Beamish10}
In particular, Beamish and coworkers demonstrated
the qualitative similarity between shear modulus and  the TO measurements.
In the shear modulus experiment solid helium is grown in between and around two closely spaced sound transducers. When one of the transducers applies an external strain, the other transducer measures the induced stress from which the shear modulus of the sample is deduced.\cite{Beamish07}
In this way, the experiment provides  a direct measurement of the elastic response to the applied force within a broad and tunable frequency range. The frequency dependence is especially crucial in determining the nature of the relaxation processes and complements current TO experiments with their limited frequency range.

Similar to the TO analysis in the previous section, we analyzed the shear modulus within the general linear response function framework, where the amplitude of the shear modulus increases (stiffens) upon lowering $T$, because of the freezing out of low-energy excitations. This change is accompanied by a prominent dissipation peak, indicative of {\em anelastic} relaxation processes.
We calculated the complex shear modulus $\mu(\omega; T)$ of a viscoelastic material and predicted: (a) the maximum of the shear modulus change and the height of the dissipation peak are independent of frequency and (b) the inverse crossover temperature $1/T_X$  vs.\  the applied frequency $\omega$  obeys the form $\omega \tau(T_X) =1$ characteristic of dynamic crossover.

\subsection{Model of dynamic shear modulus}

As in our analysis of the TO, we start with the same general linear response function formulation outlined in Section \ref{general_formalism}.
Our final result of Eq. (\ref{mu}) will, once again, reflect the general relation of Eq. (\ref{Dyson}).
Here we replace the angular coordinate of the TO with a displacement coordinate and the restoring force with a stress tensor \cite{Su10b}.
For ease, our notation in the below will differ slightly from that in Section \ref{general_formalism}.
The equation of motion for displacement $u_i$ in the $i$-th direction of a volume element in the presence of an external driving force is
\begin{eqnarray} \label{EOM}
-\rho \, \omega^2  u_i + \partial_j \,
\sigma_{ij}^{\rm He} \ \
= f_i^{\rm EXT}(\omega) + f_i^{\rm BA}(\omega) ,
\end{eqnarray}
where $\rho$ is the mass density, $f_i^{\rm EXT }$ and $f_i^{\rm BA}$ are the external force density and the backaction force density, and
$\sigma_{ij}^{\rm He}$ is the elastic stress tensor of solid helium.
In general, $\sigma_{ij}^{\rm He} = \lambda_{ijkl} \, u_{kl}$, with the elastic modulus tensor $ \lambda_{ijkl}$  \cite{LL_elastic}.
In the case of a homogeneous solid with shear wave propagation along the $z$ axis and wave polarization in the $x$-$y$ plane, the backaction takes on the form
\begin{eqnarray} \label{fBA}
f_i^{\rm BA}(\omega) = {\overline{G}}(\omega;T) \,\partial_z^2 \, u_i(\omega) ,
\end{eqnarray}
where
$\overline G$
describes the strength of the backaction on the solid (viscoelastic response) and $i=x, y$. Although $f_i^{\rm BA}(\omega)$ is much smaller than the purely elastic restoring force $\partial_j \, \sigma_{ij}^{\rm He}$, it is this term that is responsible for the stiffening of shear modulus with decreasing temperature.

Polycrystalline and amorphous materials are nearly isotropic, hence the elastic modulus tensor becomes
$\lambda_{ijkl} = \lambda_0 \delta_{ij} \delta_{kl} +
\tilde{\mu}_0 (\delta_{ik}\delta_{jl} + \delta_{il}\delta_{jk})$.
Note the stress tensor in Eq.~(\ref{EOM})  is finite only for orientations $j=z$ and either $k$ or $l$
equal  to $z$. With $k, l$ being interchangeable, the relevant element will
be $\lambda_{iziz}$, which gives the purely elastic shear modulus
$\tilde{\mu}_0$.
Finally the fully dressed shear modulus (dressed by the backaction) relates the displacement to an external force, or
$[-\rho \, \omega^2  + \mu \, \partial_z^2 ] u_i(\omega)= f_i^{\rm EXT}(\omega)$.
Comparing this expression with Eq.~(\ref{EOM}), we obtain for the dynamic shear modulus in a viscoelastic material the general response function
\begin{eqnarray} \label{mu1}
\mu(\omega; T) =
\tilde{\mu}_0(T) - \overline{G} (\omega;T).
\end{eqnarray}
Next we employ for $\overline{G}(\omega; T)$ the Cole-Cole distribution function [specifically, by reference to
Eq.~(\ref{CCD}), we set $\overline{G}(\omega;T) = g_{0} \mu_{0} G(\omega, T)$
to obtain the specific form
\begin{eqnarray} \label{mu}
\mu(\omega; T) &=& \tilde{\mu}_0 \left[
1 -\frac{g_{0}}{1-(i \omega \tau)^{\alpha}}
\right] ,
\end{eqnarray}
with the sample dependent parameter
$g_0$. The experimentally measurable quantities are the amplitude of the shear modulus, $|\mu|$, and the phase delay between
the input and read-out signal, $\phi \equiv  {\rm arg} \, (\mu)$;
$\phi$ measures the dissipation of the system, which is related to the inverse of the quality factor $Q^{-1} \equiv \tan \phi$.

Several interesting results follow immediately from the general response function in Eq.~(\ref{mu}):
(1) The change in shear modulus $\Delta \mu$ between zero and infinite relaxation time is
$\Delta \mu/\tilde{\mu}_0 = g_{0}$.
It measures the strength of the backaction as well as the concentration of defects.
(2) At fixed $T$, the shear modulus amplitude $|\mu(\omega; T)|$ decreases with increasing $\omega$.
(3) The parameter $\omega \tau$ is the only scaling parameter.
(4) The peak height $\Delta \phi$ of the phase angle is proportional to $g_{0}$. When $\omega \tau \sim1$, then
$\Delta \phi \approx g_{0} \, {\cot(\alpha \pi/4)/(2-g_{0})  } $ for $g_{0} \ll 1$.
In the limit $1 < \alpha \le 2 $, this simplifies further to
\begin{eqnarray}
\Delta \phi \approx  (1-\alpha/2)
(\Delta \mu/\tilde{\mu}_0) \ll 1 ,
\end{eqnarray}
where $\Delta \phi$ is in units of radians. Quite remarkably, the peak height $\Delta \phi$ depends only on the phenomenological parameters $\alpha$ and
$g\propto \Delta\mu$. At fixed temperature the maximum change of both $\Delta \mu$ and $\Delta \phi$ is {\it independent of frequency}.

\subsection{Results}
\begin{figure}
\begin{center}
\includegraphics[width=.75\linewidth,angle=0,keepaspectratio]{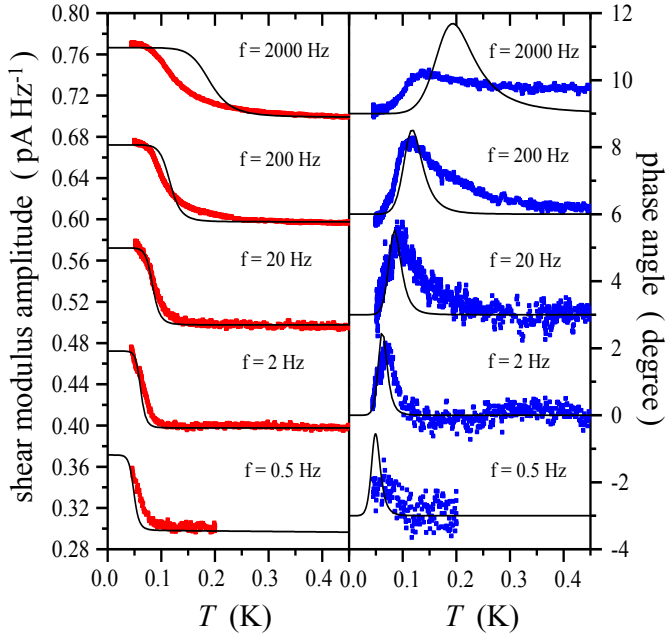}
\end{center}
\vskip-.5cm
\caption{Experimental data and theoretical calculations of the shear modulus vs.\ temperature assuming a VFT relaxation time. The red and blue squares are the experimental data for the amplitude and dissipation of shear modulus. The black-solid lines show the theoretical calculations. We used the set of parameters $\alpha=1.31$, $g_{0}=1.44\times 10^{-1}, \tilde\mu_0=0.47$ pA Hz$^{-1}$, $\tau_0=50.0$ ns, $\Delta=1.92$K,  and $T_0=-69.3$ mK. Notice a negative $T_0$ means that there is no true phase transition occurring at finite temperatures, probably because of strong quantum fluctuations of helium atoms. The shear modulus amplitude and the phase angle were shifted by 0.1 pA Hz$^{-1}$ and 3 degrees with respect to the 2 Hz data.
}\label{fig:SM_VFTnew}
\end{figure}

Let us compare our model calculations with the experimental shear modulus measurements by Beamish and coworkers \cite{Beamish10} for a transducer driven at 2000 Hz, 200 Hz, 20 Hz, 2 Hz, and 0.5 Hz \cite{Comment1}.
Specifically, for the $T$-dependent  $\tau(T)$ in  Eq.~(\ref{mu})
we consider Vogel-Fulcher-Tammann (VFT) and power-law (PL) relaxation processes.
In our model parameter search, we are not constraining $T_0$ to be positive (where $\tau$ diverges).
Fair agreement between calculations with a single set of parameters and experiments is obtained with $T_0 = -69.3$ mK,
see Fig.~\ref{fig:SM_VFTnew}. We refer to these calculations  as ``VFT$_<$''.
Similar to the TO results, a negative $T_0$ means that no real phase transition occurs in solid \HeFour.
The agreement between theory and experiment at highest (2000 Hz) and lowest (0.5 Hz) frequencies is of poorer quality.  This discrepancy may be related to very different backgrounds at these frequencies and may not be intrinsic to \HeFour. Another reason may be the presence of additional dissipation mechanisms, not accounted for.
Furthermore, our calculations confirm that the change of amplitude and  peak height of phase angle  are nearly  frequency independent between 2 Hz and 200 Hz.
By defining the crossover temperature $T_X$ as the point where the phase angle peaks, we find as predicted that $T_{X}$ decreases with decreasing $\omega$.

Finally, when we set $T_0=0$ K the VFT expression (``VFT$_0$'') reduces to an activated Arrhenius form.  While it describes reasonably well the position $T_X$,
it gives a much narrower linewidth for $\Delta\phi$  than does VFT$_<$, which is not in accord with the data and thus not shown.
Notice that the VFT relaxation is not the only possible relaxation process that can describe the data; power-law or other types of relaxation can give similar level of agreement to the experiment \cite{Beamish10}.
Iwasa proposed a relaxation process \cite{Iwasa10} similar to our phenomenological one,
which is
based on the theory of pinned vibrating dislocation lines by Granato and L\"ucke \cite{GranatoLuecke56}.
Clearly further experiments at lower frequencies and lower temperatures are required to determine the exact type of relaxation processes in solid \HeFour.

\begin{figure}[t]
\begin{center}
\parbox[t]{0.4\linewidth}{
\caption{The Cole-Cole plots for experimental data and for VFT calculation. For given form of $\tau$, all different frequency curves collapse onto one single master curve reflecting that $\omega \tau$ is the only scaling parameter. The Cole-Cole plots show reflection symmetry about Re[$|\mu-\tilde\mu_0|/\Delta \mu$]=0.5, which is a consequence of the Cole-Cole distribution function.
}\label{fig:CC}
} \hfill
\begin{minipage}{0.58\linewidth}
\includegraphics[clip, width=1.0\linewidth,angle=0,keepaspectratio]{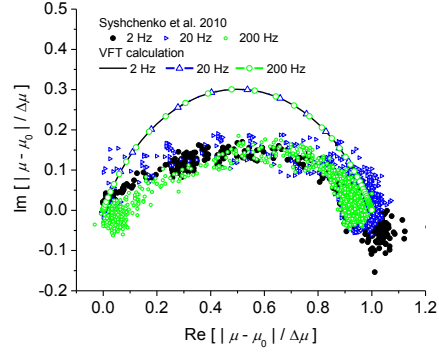}
\end{minipage}
\end{center}
\end{figure}

Figure~\ref{fig:CC} shows the Cole-Cole plot for experiments and calculations.
The experimental curves for different frequencies collapse roughly onto one curve confirming our theoretical assumption
that $\omega \tau(T)$ is a universal scaling parameter. This behavior was also seen in TO experiments \cite{Hunt09}.
In addition, the data are symmetric about
Re$[|\mu-\tilde\mu_0|/\Delta \mu]=0.5$
as expected for the Cole-Cole distribution.
A more detailed data analysis may resolve the remaining discrepancy between theory and experiment. The discrepancy may be due to either the presence of additional relaxation processes at temperatures above $T_X$, i.e., a more complicated form for $\tau(T)$, or by a modified functional form of the defect distribution function.

\begin{figure}
\begin{center}
\parbox[t]{0.4\linewidth}{
\caption{Prediction for the  inverse crossover temperature vs.\ applied frequency. The green squares correspond to $T_X$ in Ref.~\cite{Beamish10}.
For the power-law process with phase transition occurring at 40 mK, we used
$\tau=\tau_0 (|T_g|/(T-T_0))^p$ for $T>T_0$ and $\tau = \infty$ for $T \le T_0$.
}\label{fig:SM_omegavsT}
} \hfill
\begin{minipage}{0.58\linewidth}
\includegraphics[clip, width=1.0\linewidth,angle=0,keepaspectratio]{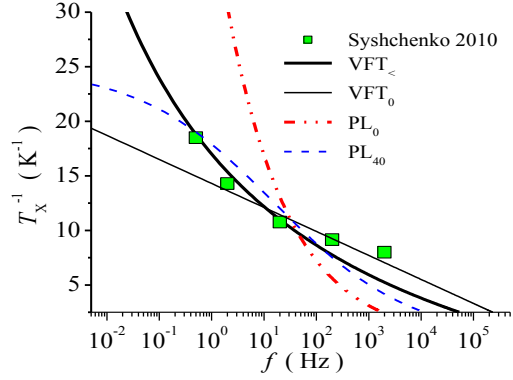}
\end{minipage}
\end{center}
\end{figure}

The pertinent question about a true phase transition at zero frequency vs.\ crossover dynamics can be addressed by investigating the asymptotic limit of $\omega \tau(T) = 1$. From this expression we estimate $T_{X}$ as a function of the applied frequency $f=\omega/2\pi$.
Figure.~\ref{fig:SM_omegavsT} shows $1/T_X$ vs.\ $f$. The VFT$_<$ and VFT$_0$ calculations give significantly better agreement than the PL calculations with phase transitions occurring at 0 K (PL$_0$) and 40 mK (PL$_{40}$). For positive $T_0$ (see PL$_{40}$), we find a true freeze-out transition, which would indicate arrested dynamics
for $f\to 0$ Hz. For both VFT and PL relaxation times our calculations demonstrate that in the low-frequency limit the existence of a phase transition shows clear signatures of $T_X$ converging toward the ideal glass temperature $T_0$. Therefore the absence of arrested behavior may serve as experimental evidence against a true phase transition.

\subsection{Viscoelastic model}

\begin{figure}
\begin{center}
\parbox[t]{0.4\linewidth}{
\caption{Lump circuit of the generalized Maxwell model. The elastic spring $\mu_0$ represents the purely elastic shear modulus at high temperature. Each Debye relaxor is made out of a series of elastic spring $\mu_{\rm RS}$ and dissipative dash-pot $\eta^{(n)}$.
}\label{fig:Maxwell}
} \hfill
\begin{minipage}{0.58\linewidth}
\includegraphics[clip, width=1.0\linewidth,angle=0,keepaspectratio]{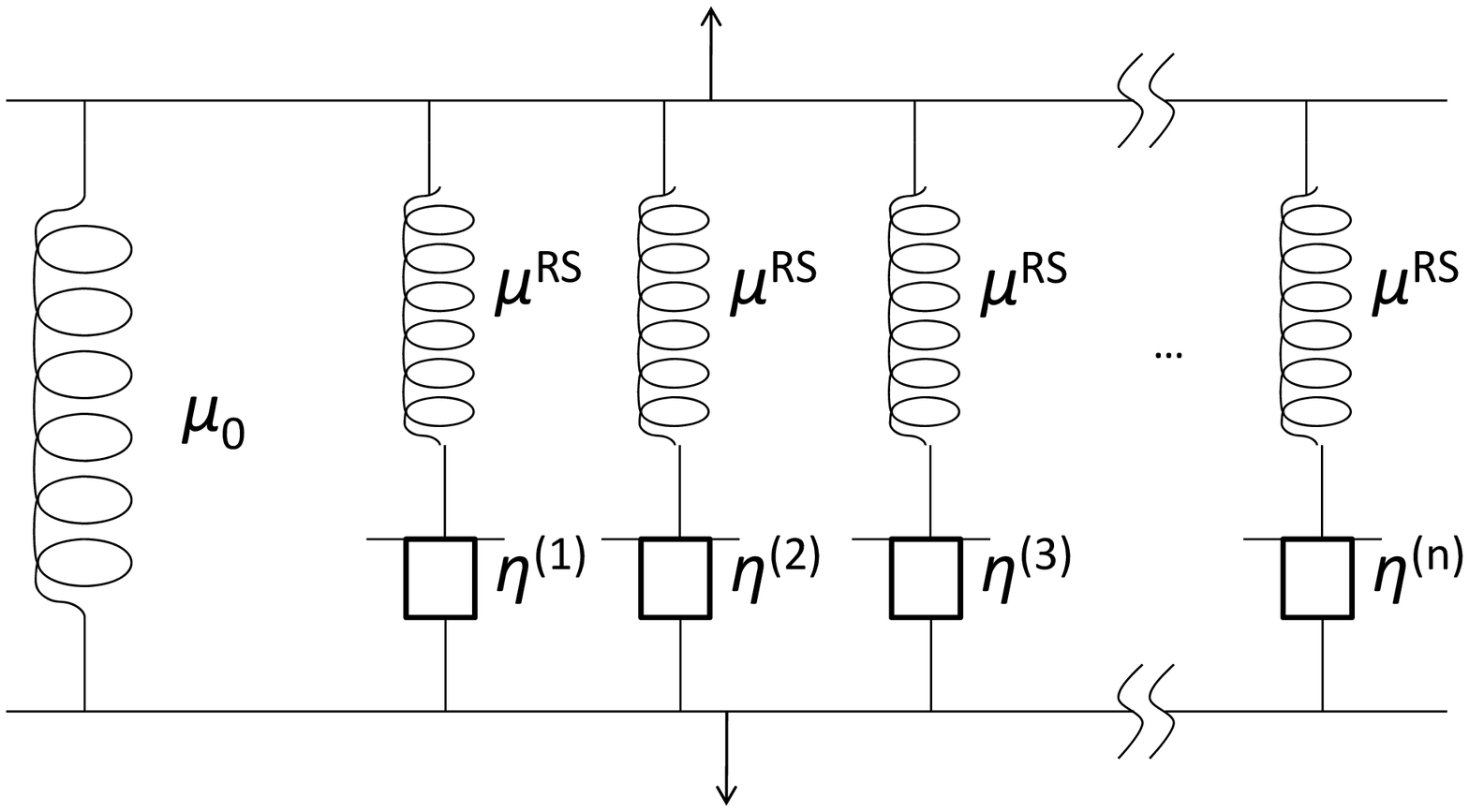}
\end{minipage}
\end{center}
\end{figure}

The viscoelastic model successfully describes
composite materials with anelastic properties. In fact, Yoo and Dorsey \cite{Dorsey08} applied the viscoelastic model to the TO experiments.
More generally, a distribution of viscous components, embedded in an otherwise elastic solid, may be treated through a generalized Maxwell model
\cite{Su10b,Su11}.
Conceptually one may subdivide the material into many elements and solve the coupled materials equations for stress and strain.
Here we use constitutive materials equations to show that our results for a glass are equivalent to the generalized Maxwell model with parallel connections of an elastic component with an infinite set of Debye relaxors, see Fig.~\ref{fig:Maxwell}.

The anomalous stiffening of the shear modulus can be
described within the viscoelastic model, though other defect mechanisms like dislocation glide are possible too \cite{Friedel,Caizhi2012}.
The equivalent lump circuit model is sketched in Fig.~\ref{fig:Maxwell}, where the basic dissipative element is the Debye relaxor. It is comprised of a serial connection of a rigid solid (RS), characterized by an elastic shear modulus $\mu_{\rm RS}$, and a Newtonian liquid (NL) or dash-pot, characterized by a viscosity $\eta$. The RS component describes the ideal elastic solid helium of this volume element, while NL represents the anelastic component, which gives rise to viscous damping. The two parts are connected in series, so that both share the same magnitude of stress, while the net strain is additive. The strain rate equation for both constituents of the Debye relaxor is
\begin{eqnarray}
\dot{\epsilon}= {\dot{\sigma}}/{\mu_{\rm RS}}
+ {\sigma}/{\eta} ,
\end{eqnarray}
where $\epsilon$ is the net strain of the Debye relaxor (DR) and $\sigma$ is the magnitude of stress shared by the components RS and NL. In order to obtain the above equation, we used the constitutive materials equations for strain and stress:
$\epsilon_{\rm RS}=\sigma_{RS}/\mu_{RS}$ and $\dot{\epsilon}_{\rm NL}=\sigma_{NL}/\eta$. After performing the Fourier transformation we obtain
for a single Debye relaxor (DR) the shear modulus $\mu_{\rm DR}=\sigma/\epsilon$,
\begin{eqnarray}
\mu_{DR}(\omega) =\frac{\mu_{\rm RS}}{1+\frac{i}{\omega \tau_{\rm DR}}}
= \mu_{\rm RS}  \left[ 1 -\frac{1}{ 1- i \omega \tau_{\rm DR} } \right] ,
\end{eqnarray}
with relaxation time $\tau_{\rm DR} \equiv \eta/\mu_{\rm RS}$. For a viscoelastic material with a single relaxation time, the  solid behaves like a parallel connection between the elastic part and the Debye relaxor
\begin{eqnarray} \label{mufinal}
\mu(\omega) &\equiv& \mu_0 + \mu_{\rm DR}(\omega) = \tilde\mu_{0} \left[
1 -\frac{g_{0}}{ 1- i \omega \tau_{\rm DR} }
\right] ,
\end{eqnarray}
with $g_{0} = \mu_{\rm RS}/\tilde\mu_0$
and  $\tilde\mu_0=\mu_0+\mu_{\rm RS}$ is the dressed elastic shear modulus.
To consider the general case of many components, we introduce Debye relaxors with different relaxation times connected in parallel as shown in Fig.~\ref{fig:Maxwell}.
The total anelastic contribution from $n$ such constituents is given by a weighted sum. The corresponding continuous version of this expression with a distribution $P(s)$ of relaxation times is
\begin{eqnarray} \label{GMG}
\mu_{ae}(\omega) = \mu_{\rm RS} \int_0^{\infty}
ds \ P(s)
\,\left[1- \frac{1}{1- i \omega \tau s} \right]
=\mu_{\rm RS} - \frac{\mu_{\rm RS}}{ 1-(i \omega \, \tau) ^{\alpha} } .
\end{eqnarray}
Here the Cole-Cole distribution,
\begin{eqnarray}
P(s)=\frac{t^{-(1-\alpha)} \sin \alpha \pi  }
{1+s^{2 \alpha}+2 s^{\alpha}\cos \alpha \pi} ,
\end{eqnarray}
was used.
Similar to our response functions elsewhere, the net shear modulus of the composite system is given by the sum of two terms- the purely elastic component and the anelastic contribution
\begin{eqnarray} \label{mufinal}
\mu(\omega) &=& \mu_0 + \mu_{ae}(\omega) =  \tilde\mu_{0} \left[
1 -\frac{g_{0}}{ 1-(i \omega \tau)^{\alpha} }
\right] .
\end{eqnarray}
Indeed, this expression is identical to the one obtained previously using the general response function formalism with a backaction (\ref{Dyson}).
This is no coincidence, since the backaction term accounts for damping and thus describes the anelastic response of defects to the external stress.


\section{Dielectric properties of a viscoelastic medium}
\label{diels}

The measurements of $\epsilon(\omega, T)$ by Yin et al.\ \cite{Yin11} showed the anomalous increase of the dielectric function of solid \HeFour\
at low temperatures. A similar test experiment in liquid helium showed no such effect. We propose that these results may be explained by an electro-elastic coupling of a quenched solid with frozen-in internal stress.
Such behavior cannot be described by the standard Clausius-Mossotti equation via a change in mass density or polarizability (due to, e.g., dipole-induced dipole interactions).
Neither the measured change of the mean mass density $\delta \rho/\rho \sim 10^{-6}$,
nor the predicted correction in polarizability, which actually leads to a decrease of $\epsilon(\omega, T)$ at low temperatures \cite{Kirkwood36, Chan77},
can account for the reported anomalous change of the dielectric function of order
$\delta \epsilon /\epsilon \sim 10^{-5}$, while a model with frozen-in stress and electro-elastic coupling can explain the data.

\subsection{Model for electro-elastic properties}

The relation of Eq. (\ref{el}) that we derive below constitutes yet another realization of our general relation of Eq. (\ref{Dyson}). We now turn to the specifics of the electro-elastic coupling
describing the interaction between the electromagnetic and the strain fields. The coupling may be obtained by expanding the dipole moment, $\bf p (r_{\it a})$, of a nonpolar atom around its equilibrium value:
\begin{eqnarray}\label{TEdipole}
{\bf p (r_{\it a}) \approx p (R_{\it a})+ (u_{\it a} \cdot \nabla ) \ p \left.  \right|_{R_{\it a}} } ,
\end{eqnarray}
where $\bf r_{\it a}=R_{\it a}+u_{\it a}$ is the position of atom $a$, $\bf R_{\it a}$ is its equilibrium position, and $\bf u_{\it a}$ is its displacement. The polarization is obtained by averaging ${\bf p}$ over a macroscopic volume element $v$,
${\bf P (r)}=({1}/{v}) \sum_{\it a} {\bf p (r_a)}$. In the continuum limit (when the macroscopic length scale is far larger than the atomic length)
$\sum_a {\bf p (r_{\it a})}  \approx (1/v)\int_v d {\bf r}' \ {\bf p (r')}$ .
In the presence of a local strain field, to linear order in the displacement, Eq.~(\ref{TEdipole}) reads
\begin{eqnarray}
 {\bf P (r)} \approx (1/v^2)\int_v d \,{\bf R \ [ \,p(R)
+ (u \cdot \nabla) \ p (R)\,]}.
\end{eqnarray}
 Integration of the first term yields the macroscopic polarization for zero internal strain (a solid in equilibrium), ${\bf P}_0$. It is related to the macroscopic electric field ${\bf E}$ by ${\bf P}_0 \equiv \chi_0 {\bf E}$ where
$\chi_0 = (\epsilon_0-1)/4\pi$ is the zero-strain susceptibility and $\epsilon_0$ is the permittivity.  The second term in the integration describes the polarization change $\bf \delta P$ due to atomic displacements.
Neglecting surface contributions the second term modifies the polarization ${\bf P}={\bf P}_0 + \delta {\bf P}$ by
\begin{eqnarray} \label{P-strain}
\delta {\bf P} = - (1/v^2) \int d {\bf  R \  ( \nabla \cdot u )  \ p(R)} \approx - e_{ii} \,{\bf P}_0 ,
\end{eqnarray}
with $e_{ii} = (1/v) \int_v d{\bf R} \, (\nabla \cdot \bf u)$ the macroscopic frozen-in dilatory strain (we use the repeated indices summation convention
$e_{ii}\equiv e_{11} + e_{22} + e_{33}$). In obtaining Eq.~(\ref{P-strain}), we assumed that ${\bf P}$ is slowly varying. This long-wave length approximation holds  for wavelengths of microwave ($\lambda >10^{-3}$m) and above. The polarization change alters the dielectric function $\epsilon_{ii}=(\epsilon_0)_{ii} + \delta\epsilon_{ii}$ by
\begin{eqnarray} \label{deltaepsilon}
\delta \epsilon_{ii} \, (\omega, T)
= - 4 \pi \chi_0 \, e_{ii} \, (\omega, T)
\end{eqnarray}
with $\epsilon_{ii} -1 = 4 \pi P_i/ E_i$. Only the diagonal components of the strain tensor play a role in a leading order expansion of the electro-elastic coupling.
In principle, the shear strain can couple to the electric field by considering dipole-induced dipole interactions (van der Waals), which is
a higher order effect. The same coupling mechanism between photon (electric field) and phonons (strain) gives rise to the acousto-optical effect.
Our derivation of Eq.~(\ref{deltaepsilon}) is, to leading order, equivalent to the Pockels coefficient for acousto-optical coupling in
isotropic dielectrics,
$
\delta \epsilon_{ij}= - \epsilon^2 \,[ 2 P_{44} \, e_{ij} + P_{12} \, e_{kk} \delta_{ij}],
$
where $P_{kl}$ are the reduced Pockels coefficients (of order unity) \cite{Werthamer69, Kisliuk91, Landau_Lifshitz}.

We next turn to the locally frozen-in strains and write the equation of motion within the general response function theory.
As before in an isotropic medium the elastic stress tensor
 $\sigma_{ij}^{\rm He} = \lambda_{ijkl} \, \partial u_k/\partial x_{l}$
with $\lambda_{ijkl} = \lambda_0 \delta_{ij} \delta_{kl} + \mu_0( \delta_{ik}\delta_{jl} + \delta_{il}\delta_{jk})$.
If the electric field  couples  to local density fluctuations only through dilatory strain,
then the important matrix element is the Lam\'e parameter $\lambda_{0}$. We write the displacement to an out-of-equilibrium internal (INT) force in the presence of the backaction as
\begin{eqnarray} \label{eEOM}
-\rho  \omega^2  \,u_i (\omega)+ \lambda_{0} \, \partial_i^2 u_i (\omega)
= f_i^{\rm INT} (\omega)+ f_i^{\rm BA} (\omega),
\end{eqnarray}
where $u_i$ is a displacement in the $i$th  direction and
$\rho$ is the mass density.
The backaction force density of nearby atoms is given by
\begin{eqnarray} \label{fBA}
f_i^{\rm BA} (\omega) = \overline{G}(\omega;T) \,\partial_i^2 \, u_i(\omega) ,
\end{eqnarray}
where ${\overline{G}}$ is the strength of the pertinent backaction.
$f_i^{\rm INT }$ is the out-of-equilibrium (frozen-in) internal force density in the $i$th direction at the defect.
Integrating Eq.~(\ref{eEOM}) yields
the strain due to the internal dilatory stress $\sigma_{ii}$:
\begin{eqnarray}
{e}_{ii}(\omega, T) = {\sigma}^{}_{ii}
\left(\lambda_{0}-\overline{G} (\omega; T) \right)^{-1}.
\end{eqnarray}
Again, we assume that the backaction can be described by a Cole-Cole distribution of Debye relaxors,
$\overline{G}(\omega; T) =  {g_0} \lambda_0/{\big[1-(i \omega \, \tau) ^{\alpha}\big]}$.
The corresponding local dilatory strain reads
\begin{eqnarray} \label{strain_glass}
{e}_{ii} (\omega, T) &=& e_0 \left(1- {g_0}/\big[{1-(i\omega \tau)^{\alpha}}\big] \right)^{-1} ,
\end{eqnarray}
where $e_0 \equiv \sigma_{ii}/\lambda_0$
at $T=0$.
From Eqns.~(\ref{deltaepsilon}) and( \ref{strain_glass})  the change in the dielectric function due to local strain fluctuations is
\begin{eqnarray}
\label{el}
\delta \epsilon_{ii} &=& -4 \pi \chi_0 \, e_0
\left(1- {g_0}/\big[{1-(i\omega \tau)^{\alpha}}\big] \right)^{-1} .
\end{eqnarray}
This result is similar to the one for the TO and shear modulus discussed in the earlier sections. At low temperatures, $\tau \rightarrow \infty$ and $e_{ii} \rightarrow e_0$, hence the strain is minimal and the reduction of  the dielectric function due to local strain fluctuations is small.
At high temperatures, $\tau \rightarrow 0$ and $e_{ii} \rightarrow e_0 (1-g_0)^{-1}$ reaches its maximum resulting in the largest reduction of  strain,  where solid $^4$He is softest.
The main result is that  the dielectric function reflects the arrested dynamics of the glassy components at low temperatures through the acousto-optical (electro-elastic) coupling.

The  derivation of Eq.~(\ref{el}) for the dielectric response is extremely general.
We illustrate how a Cole-Cole form for the elastic relaxation implies a similar dielectric response (and vice versa). An identical result holds for other forms of the local elastic relaxation dynamics- these will leave a similar imprint on the dielectric response.
Historically (since the 1940s) the Cole-Cole response function \cite{Phase1} was found to be valid
for the dielectric response. In our initial works on \HeFour\, in trying to capture the local quenched dynamics, we first assumed this form for the TO and then for the general elastic response.
By virtue of their inter-relation and coupling, the local relaxation dynamics is the same for all of these quantities. Therefore measurements of the dielectric response may inform
about local dynamics and  vice versa. Practically, our prediction of Eq.~(\ref{el}) applies to any nonpolarized system with a local distribution of stress relaxations.
In polarized materials, detecting a change in the dielectric function due to atoms sensing different local fields is challenging because the large intrinsic polarizability of the material will overwhelm the contributions derived above.
Among the nonpolarized solids, solid $^4$He favors the observation of this phenomenon especially because of its softness. Rapid cooling of solid helium allows a large local strain build-up, which is proportional to the size of the effect. Similarly, we expect the effect to be seen via delicate effects in solid $^3$He, hydrogen or xenon.

\subsection{Results}

\begin{figure}
\begin{center}
\parbox[t]{0.4\linewidth}{
\caption{Experimental data and calculations of the dielectric function vs.\ temperature.
The red circles are the experimental data of the dielectric function (data by Yin et al.\ \cite{Yin11}).
The black lines are the calculated amplitude and phase lag (dissipation) of $\epsilon(\omega; T)$.
We used parameters $\alpha=1.49$, $e_0=8.88\times 10^{-4}, g=0.21$, $\tau_0=10.4$ ns, $\Delta=1.92$ K, and $T_0=-119$ mK.
}\label{fig:epsilon}
} \hfill
\begin{minipage}{0.58\linewidth}
\includegraphics[clip,width=1.0\linewidth,angle=0,keepaspectratio]{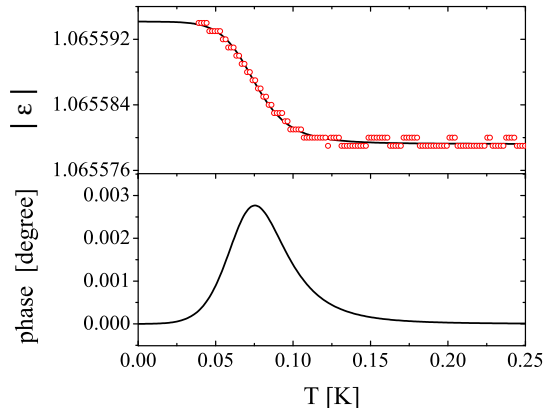}
\end{minipage}
\end{center}
\end{figure}

The results of our electro-elastic predictions for the dielectric function are shown in Fig.~\ref{fig:epsilon}.
We obtain excellent agreement with experiment  \cite{Yin11} for an applied alternating voltage at 500 Hz.
Our analysis predicts, that similar to the TO and shear modulus, a dissipation peak appears in the dielectric function.
The phase lag angle
$\phi=arg( \epsilon )$ records the lag between the real and imaginary part of $\epsilon(\omega; T)$.
Future observation of the dissipation peak will provide an important test of our picture concerning
quenched dynamics in solid \HeFour.
Consistent with earlier sections in this review, we assumed a Vogel-Fulcher-Tammann (VFT) form for the defect relaxation time $\tau(T) = \tau_0 \,e^{\Delta/(T-T_0)}$.
As in the previous sections, we obtain from our fits a negative $T_0=-119$ mK, in accordance with an avoided dynamic arrest of defect motion.

A rough estimate of the electro-elastic coupling can be obtained from mass flow measurements in bulk solid $^4$He, \cite{Ray10} where a pressure difference of $\Delta P_{L} \sim 0.1$ bar across a centimeter-sized pressure cell was reported. The estimated local strain, with a bulk modulus $B=320$ bar, is accordingly  $\Delta P_{L}/B = 3 \times 10^{-4}$. This is consistent with the value we used for the fit in Fig.~\ref{fig:epsilon},  namely $e_0 = 8.88 \times 10^{-4}$.
In addition, the $P(T)$  measurement by Yin  clearly deviates from a purely Debye lattice behavior at around $T = 0.4$ K with a large positive intercept corresponding to the order of 100 {\em ppm} of TLS, see Fig.~\ref{fig:PT}. This number is roughly five times larger than the most disordered sample in Lin's \cite{Lin09} specific heat experiments on ultrapure $^4$He with less than 1 ppb of $^3$He impurities \cite{Lin09,Su10}.
Thus the crystals grown by Yin are strongly disordered and harbor sufficiently many defects to support
centers of local strain fields.

Thus far, we assumed that both local and global stress are constant at low temperatures.
From Fig.~\ref{fig:PT} we can read off that the global pressure change between 300 mK and 40 mK is less than $\Delta P_T=0.18$ mbar. This is more than three orders of magnitude smaller than the local stress $\sigma_L = 8.88 \times 10^{-4} \times 320 \ {\rm bar} \sim 250$ mbar inferred from the dilatory strain $e_0$ used in the fit, as well as the static pressure difference  $\Delta P_L = 0.1$ bar measured at two pressure gauges in  mass flow experiments \cite{Ray10}.
Putting all  the pieces together, we find that the change of the dielectric function based on global
density changes in the Clausius-Mossotti equation is negligible.
This is because  the corresponding density change,
$\Delta\rho/\rho = \Delta P_T/B < 10^{-6}$, leads to a change in the
 dielectric function of only
$\delta \epsilon \approx (\epsilon-1) \Delta \rho /\rho = 0.065 \Delta P_T/B < 10^{-7}$, which is more than two orders of magnitude too small to account for the observed effect of order $10^{-5}$.
Whereas the model of local stress and electro-elastic coupling can  account for the magnitude and temperature dependence of the dynamic dielectric function.
 More recently Yin {\it et al.} \cite{Yin2012} redesigned their dielectric function experiment with a simplified capacitor geometry
and found no measurable anomaly at low temperatures.
Within our theory of disorder such a null result is consistent with a negligible amount of frozen-in stress in the solid.


\section{Conclusions and future directions}

In this review we provided a general overview of the role of defects in solid \HeFour. We suggested that defect dynamics and freeze-out provides a rich ground to account for a significant fraction of the data, while at the same time they allow enough flexibility to accomodate sample and history dependence and hysteretic behavior ubiquitously seen in experiments.    The  general response function approach presented covers a wide range of observed effects.
We provide a brief synopsis of these below.

\subsection{Thermodynamics}

 We started our discussion with the notion that  any true phase transition, including supersolid,  is accompanied by a thermodynamic signature. Therefore   thermodynamic measurements like specific heat can reveal the signature one would naturally associate with a phase transition. However, we found that the  excess specific heat and corresponding entropy  are  consistent with the contribution from noninteracting two level systems (TLSs). The estimated fraction of these TLS to the specific heat is at the level of tens to hundreds of parts-per-million. Consequently, the corresponding entropy contribution  $\Delta S \sim 10^{-4}$ J/(K mol)  is inconsistent by
several
orders of magnitude with reported values of the superfluid fraction
(NCRIF)
in the TO and shear modulus experiments. We also point out that some of the most disordered samples demonstrated "supersolid" fraction up to 20\%, while there is little evidence for any excess entropy on the scale of the gas constant $R$ times the fraction of defects in any measured sample.

\subsection{Single and double torsional oscillators}
 We have shown that a phenomenological glass (viscoelastic) model for quenched defects accounts for the experimentally observed change in resonant frequency and the concomitant peak in dissipation.
Our analysis of torsional oscillator (TO) experiments revealed that most are well described by a Cole-Cole
distribution for  relaxation times.
In addition, we derived a simple relation for the ratio of change in dissipation
and change in resonant frequency that can explain the large ratios  observed
in experiments.
The values for the glass exponents in the distribution function of the backaction required to fit the experimental data
point toward broad distributions of glassy relaxation times, which clearly invalidate any attempt to describe these
experiments by a single overdamped mode, i.e., a single Debye relaxation process.

We also  applied these ideas to understand the double oscillator data.
Our studies of the coupled oscillator showed that the observed shifts in resonant frequencies and dissipation are in agreement
with a glassy backaction contribution  provided  one includes anomalous damping in the dummy bob and an explicit frequency dependence in the backaction term.
As a side comment, it came as a surprise that already the unloaded double TO  (no \HeFour) required a negative
damping coefficient for the dummy bob to accurately describe resonant frequencies and dissipation at 300 mK.

Finally, one should keep in mind that a significant difference between glassy and supersolid dynamics is that a glassy contribution to the TO grows with frequency, while a superfluid component decreases with frequency. This could be another differentiating factor for separating very different relaxation mechanisms.

\subsection{Shear modulus}

We showed that  the shear modulus anomaly of solid $^4$He is strongly affected by the dynamics of defects. The freezing out of defects leads to stiffening of the solid concomitant with a peak in dissipation.  By studying the glass susceptibility due to the backaction, we found that both the amplitude change and $T$-dependence of the shear modulus are well captured by this model.
An important consequence of the dynamic response analysis was the description of the dissipation or phase angle.
We found that the peak height of the dissipation is independent of the applied frequency and linearly proportional to the Cole-Cole exponent $\alpha$ as well as the backaction strength $g_{0}$. As $g_{0}$ depends on the concentration of the TLS, we predicted that increasing disorder will result in larger amplitude changes of the shear modulus.
Additionally, we extracted a universal scaling behavior proportional to $\omega \tau(T)$ using the Cole-Cole plot.
In this plot all curves of the shear modulus collapsed onto a single curve over a wide range of frequencies.
Furthermore, we have shown that
the glass contribution can be described by a viscoelastic model through the incorporation of anelastic elements
in constitutive equations of stress and strain.

\subsection{Dielectric function}
We have shown that the arrested glass dynamics causes the low-temperature anomaly in strained solid $^4$He through the acousto-optical (electro-elastic) coupling and proposed that the temperature behavior of the dielectric function is coupled  to local strain fields near crystal defects.
It records the glassy dynamics and  freeze-out of the hypothesized TLS excitations, which also lead to a stiffening of the solid with decreasing temperature. This effect is not captured by the standard Clausius-Mossotti relation, which attributes dielectric function changes to a change in mass density or polarizability of the nonpolar \HeFour\ atom.
An important consequence of the phenomenological glass susceptibility is the decrease of the dielectric function at high temperatures, accompanied by a broad dissipation peak that  can be measured  by the imaginary part of the dielectric function. We hypothesized that the cooperative motion of atoms forming the TLS along dislocation segments is the relevant process contributing to the reported anomaly.
In our model, both the change in $\epsilon(\omega; T)$ and the dissipation peak are to leading order independent of the applied frequency.
Since  the coefficient $g_0$ of the backaction depends on the concentration of defects, we predicted that the change in dielectric function will be larger in quench-cooled or shear-stressed samples, while it vanishes in defect-free single crystals. Beyond the specific application to solid \HeFour,
which we invoked here,
our formalism allowed for a direct link between elastic and dielectric properties which could prove
fruitful in many other systems.


\bigskip

In summary we encourage and welcome experiments that will allow a more precise structural characterization of \HeFour. Also one needs to sharpen and provide a more detailed analysis of the structural aspects of  solid \HeFour\ in order to be able to investigate any arrested dynamics of the postulated glassy regions to separate it from a simple crossover phenomenon.
Finally, we believe that more dynamic studies probing the frequency or time response to a stimulus are necessary to
investigate the differences between small subsystems of glassy, supersolid or superglassy origin in solid \HeFour.

\begin{acknowledgements}
We are grateful to colleagues and collaborators who provided encouragement and constructive criticism of the ideas presented here, A.F. Andreev, P. W. Anderson,  I. Beyerlein, J. C. Davis, A. Dorsey,  C. Reichardt, B. Hunt, E. Pratt, V. Gadagkar,  J. Reppy, M. Chan, N. Prokof'ev, B. Svistunov, D. Schmeltzer, A. Kuklov and E. Rudavsky.  This work was supported by the US Dept.\ of Energy at Los Alamos National Laboratory
under contract No.~DE-AC52-06NA25396 and the Basic Energy Sciences Office.
Z. N. was partially supported by the NSF grant Award No.\ DMR-1106293. We also acknowledge  steady and generous support by  the Aspen Center for Physics where parts of this review were written.

\end{acknowledgements}

\end{document}